\documentclass[aip,preprint]{revtex4-1}
\usepackage{graphicx}

\usepackage{appendix}
\usepackage{amsmath}
\usepackage{comment}
\usepackage{color}

\draft 

\begin{document}

\title{A sign-blocking method for mitigating the fermion sign problem}

\author{Yunuo Xiong}
\email{xiongyunuo@hbpu.edu.cn}
\affiliation{Center for Fundamental Physics, Hubei Polytechnic University, Huangshi 435003, China}

\author{Hongwei Xiong}
\email{xionghongwei@hbpu.edu.cn}
\affiliation{Center for Fundamental Physics, Hubei Polytechnic University, Huangshi 435003, China}
\date{\today}

\begin{abstract}
The fermion sign problem remains the primary obstacle in simulating the thermodynamic properties of various fermionic systems. In this work, we present a sign-blocking method to mitigate the numerical instability inherent in the sign problem. In the sign-blocking method, the Monte Carlo importance sampling remains identical to traditional methods; instead, the sign-blocking method is applied during the post-processing of signed samples. Given the significant progress in simulating the 2D Fermi-Hubbard model over the past decade, a wealth of energy benchmarks is available for comparison. Consequently, we use the 2D Fermi-Hubbard model as a benchmark to validate the sign-blocking method. Surprisingly, our results align exceptionally well with existing state-of-the-art benchmarks, even in regimes previously considered challenging.
The physical mechanism of the sign-blocking method lies in uncovering the correlation between energy and sign factors through data blocking, thereby successfully inferring the fermionic system's energy.
Our findings suggest that the sign-blocking method holds promise for complex quantum systems, particularly when combined with appropriate simulation techniques such as auxiliary-field formalisms that trace out the fermionic degrees of freedom.
\end{abstract}
\maketitle

\section{Introduction}

The fermion sign problem \cite{loh1990sign,dornheim2019fermion,troyer2005computational,alexandru2022complex,dornheim2018uniform,he2025revisiting} remains the primary obstacle to simulating the thermodynamic properties of many-body fermionic systems. In the path-integral framework, the integrand's weight often oscillates between positive and negative values, which prevents the definition of a positive-definite probability distribution required for Monte Carlo importance sampling.
While the standard reweighting method \cite{dornheim2019fermion} attempts to bypass this by using the absolute value of the weights to construct estimators, the resulting signal-to-noise ratio decays exponentially. Consequently, the exponential scaling of computational cost with increasing system size or decreasing temperature constitutes the widely recognized exponential computational bottleneck.

Over the past decade, a variety of developments \cite{wu2005sufficient,prokof2007bold,booth2009fermion,schoof2011configuration,brown2013path,li2016majorana,dornheim2015permutation,hirshberg2020path,dornheim2020attenuating,dornheim2018uniform, vitenburgs2026, bonitz2024toward,vorberger2025roadmap,bonitz2026} have advanced our ability to address the sign problem. For instance, CP-AFQMC (constrained path auxiliary-field quantum Monte Carlo)\cite{zhang1997constrained} uses a trial wavefunction to restrict the configuration space to regions where sign fluctuations are effectively removed. The fictitious identical particles method \cite{xiong2022thermodynamic,xiong2023thermodynamics,dornheim2023fermionic,dornheim2025unraveling,dornheim2025taylor,morresi2025normal,fan2025quantum} introduces a continuous parameter $\xi$ to perform simulations in sign-problem-free bosonic sector ($\xi \geq 0$), subsequently extrapolating these results to the fermionic limit ($\xi=-1$). The pseudo-fermion method \cite{xiong2025pseudo,YunuoUEG} approximates fermionic behavior by treating the number of imaginary-time slices $M$ as a variational parameter while sampling the absolute value of the fermion propagator. In this work, we propose a fundamentally different approach. Rather than imposing constraints on the sampling space or extrapolating from the bosonic sector, we mitigate the sign problem by directly extracting the intrinsic correlation between the energy and its corresponding sign factor.

Due to the fermion sign problem, configurations generated via importance sampling in Monte Carlo simulations are characterized by both positive and negative weights. As demonstrated by Dornheim \cite{dornheim2019fermion}, extracting fermionic properties from the sign distribution in isolation is an inherently formidable task. The primary limitation of such an isolated analysis is its inability to capture the critical information arising from the interference between positive and negative weight sectors. Although the standard reweighting method \cite{dornheim2019fermion} formally retains this interference, the underlying correlation information vanishes if the sign and energy are processed as independent statistical distributions. Since the energy and sign of a given configuration are intrinsically correlated, merely examining their marginal distributions cannot yield the true fermionic energy. However, by extracting the underlying correlation between these variables, it becomes possible, in principle, to reconstruct the fundamental physics of the fermionic system.

To extract the correlation between energy and sign from the cumulative interference of weights while maintaining computational feasibility, we partition the ensemble of signed samples into $\mathcal{F}$ blocks of size $K$. Within each block, samples of opposing signs coexist; by constructing a block-wise estimator that explicitly incorporates these sign factors, we preserve a portion of interference required to capture fermionic physics. To mitigate the numerical instability inherent in the sign problem, we utilize the absolute value of the mean for each block as the primary estimator. We refer to this procedure as the sign-blocking method, as it leverages statistical blocking to harvest the sign-energy correlation.

To demonstrate the validity of this concept, we employ the two-dimensional (2D) Fermi-Hubbard model as a testbed, leveraging well-characterized benchmarks established over the past decade to cross-validate our approach. Our implementation utilizes the determinant quantum Monte Carlo (DQMC) framework \cite{blankenbecler1981monte, hirsch1985two, lin1987two, hirsch1988pairing, white1989numerical, sorella1989novel, loh1990sign, santos2003introduction, huang2017numerical, assaad2025alf, song2025extended}, employing an auxiliary-field transformation to deal with interactions, which can be regarded as a special case of AFQMC applied to the Fermi-Hubbard model. Similar to DQMC, AFQMC is generally subject to the fermion sign problem. A standard remedy is the constrained-path AFQMC \cite{zhang1995constrained}, which adapts fixed-node method \cite{anderson1975random,Ceperley1980} to restrict the sampling space to circumvent the fermion sign problem. For several representative parameters, the energy per site inferred by the sign-blocking method is in excellent agreement with state-of-the-art methodologies \cite{leblanc2015solutions, zheng2017stripe}. This agreement offers a promising perspective for reconciling disparate results obtained through various numerical techniques. Notably, by maintaining an importance sampling procedure identical to existing approaches while differing only in the post-sampling data processing, the sign-blocking method provides a versatile framework for identifying suitable fermionic systems and enhancing simulation methods in future investigations.

The paper is organized as follows. In Sec. \ref{SigIntr}, we introduce the formal framework of the sign-blocking method, detailing how it captures the statistical interference between positive and negative weight regions. In Sec. \ref{Appl}, as an example, we describe the practical implementation for the 2D Fermi-Hubbard model, including the formulation of the scaling \textit{ansatz} for the optimal block size. In Sec. \ref{Results}, we present our numerical results for several representative cases of the 2D Fermi-Hubbard model. We benchmark our predictions against other state-of-the-art methods, including FN, DMRG, DMET, iPEPS and CP-AFQMC\cite{leblanc2015solutions,zheng2017stripe}, \textit{etc}. In Sec. \ref{Strategy}, we examine the sign-blocking method within the context of fermionic propagator-based Monte Carlo sampling, and outline a strategy for wider application. In Sec. \ref{correlation}, we discuss the mechanism of the sign-blocking method as a correlation extraction technique. Finally, we provide a summary and discussion of the sign-blocking method and future outlook in Sec. \ref{Summary}.

\section{The sign-blocking method to mitigate the fermion sign problem}
\label{SigIntr}

The partition function of a fermionic system can be expressed as:
\begin{equation}
    Z=\text{Tr}(e^{-\beta\hat H})=\sum_{\mathcal{C}}W(\mathcal C).
\end{equation}
Here, $\beta$ denotes the inverse temperature, defined as $\beta = 1 / (k_B T)$ with $k_B$ the Boltzmann constant. In the context of quantum Monte Carlo simulations, $\beta$ represents the total length of the imaginary-time evolution. As $\beta \to \infty$, the system's properties converge toward those of the ground state.
$\mathcal{C}$ denotes all possible configurations in the sampled space. Unfortunately, for most fermionic systems, the exchange antisymmetry of the many-body wavefunction results in weights $W(\mathcal{C})$ that can be either positive or negative. This fundamental feature is the origin of the notorious fermion sign problem. We can concisely define $G$ signed samples as follows:
\begin{equation}
    \{S_1,\mathcal{C}_1\},\cdots,\{S_j,\mathcal{C}_j\}\cdots,\{S_G,\mathcal{C}_G\}.
    \label{signG}
\end{equation}
Here, the sign factor $S_j$ is either $+1$ or $-1$.

To handle these non-positive weights in stochastic simulations, a standard approach is to perform reweighting. In this scheme, importance sampling is conducted using the absolute weights $|W(\mathcal{C})|$, and the expectation value of a physical observable $\hat{O}$ is calculated as:
\begin{equation}
\langle \hat{O} \rangle = \frac{\sum_{\mathcal{C}} O(\mathcal{C}) W(\mathcal{C})}{\sum_{\mathcal{C}} W(\mathcal{C})} = \frac{\sum_{\mathcal{C}} O(\mathcal{C})S(\mathcal{C})|W(\mathcal{C})|}{\sum_{\mathcal{C}} S(\mathcal{C})|W(\mathcal{C})|}=\frac{\langle O \cdot S \rangle_{|W|}}{\langle S \rangle_{|W|}},
\end{equation}
where $S(\mathcal{C}) = W(\mathcal{C})/|W(\mathcal{C})|$ represents the sign of the configuration $\mathcal{C}$. $\langle\cdots\rangle_{|W|}$ represents the mean relative to the positive weight $|W|$. While mathematically exact, this reweighting procedure leads to an exponential decay of the average sign $\langle S \rangle_{|W|}$ with respect to the system size and inverse temperature, causing the statistical relative error to explode for large system at low temperature.

Once a total of $G$ de-correlated signed samples $\mathcal{C}_j$ (where $j=1, \dots, G$) are generated via importance sampling according to the probability $P(\mathcal{C}_j) \propto |W(\mathcal{C}_j)|$, the expectation value of an observable $\hat{O}$ is estimated as a ratio of two sample averages:
\begin{equation}
\langle \hat{O} \rangle = \frac{\frac{1}{G} \sum_{j=1}^{G} O_j \cdot S_j}{\frac{1}{G} \sum_{j=1}^{G} S_j} = \frac{\bar{O}_{\text{sign}}}{\bar{S}}.
\label{sampleS}
\end{equation}
In this framework, $\bar{S}$ represents the average sign.
Here $S_j$ represents the sign of the $j$-th sample, as shown in Eq. (\ref{signG}). In principle, the exact result is recovered in the limit of infinite $G$, such that $\lim_{G \to \infty} \langle \hat{O} \rangle = O_{\text{exact}}$.

From Eq. (\ref{sampleS}), we observe that the samples with positive signs and those with negative signs undergo a form of statistical interference. If we were to forcibly ignore this interference by taking the absolute value of the weights, we would obtain the following estimator:
\begin{equation}
{O}_{\text{ignore}} = \frac{\frac{1}{G} \sum_{i=1}^{G} O_j \cdot |S_i|}{\frac{1}{G} \sum_{i=1}^{G} |S_i|}.
\label{k1}
\end{equation}
However, usually this approach fails to provide an accurate prediction of the system's true physical properties, as it neglects the “interference” between positive and negative weights.

In general, $O_{\text{ignore}}$ exhibits a non-negligible deviation from the true fermionic properties. However, for homogeneous systems, a straightforward yet improved approximation can be implemented. We assume that, for a small system, the exact result $O_{\text{exact}}$ can be obtained using methods such as exact diagonalization (ED). We then define the ratio between these two quantities as:
\begin{equation}
    r=\frac{O_{\text{exact}}}{O_{\text{ignore}}}.
\end{equation}
For a larger homogeneous system, once $O_{\text{ignore}}$ is obtained, we may consider to approximate the true fermionic property as:
\begin{equation}
    O_F\approx r \cdot O_{\text{ignore}}.
    \label{simple}
\end{equation}
As will be demonstrated later, this simple calibration scheme yields energy that even outperforms the fixed-node method for a typical Fermi-Hubbard model. This finding suggests that further development of more sophisticated approximation methods is a promising avenue.

\begin{figure}[htbp]
\begin{center}
\includegraphics[scale=0.4]{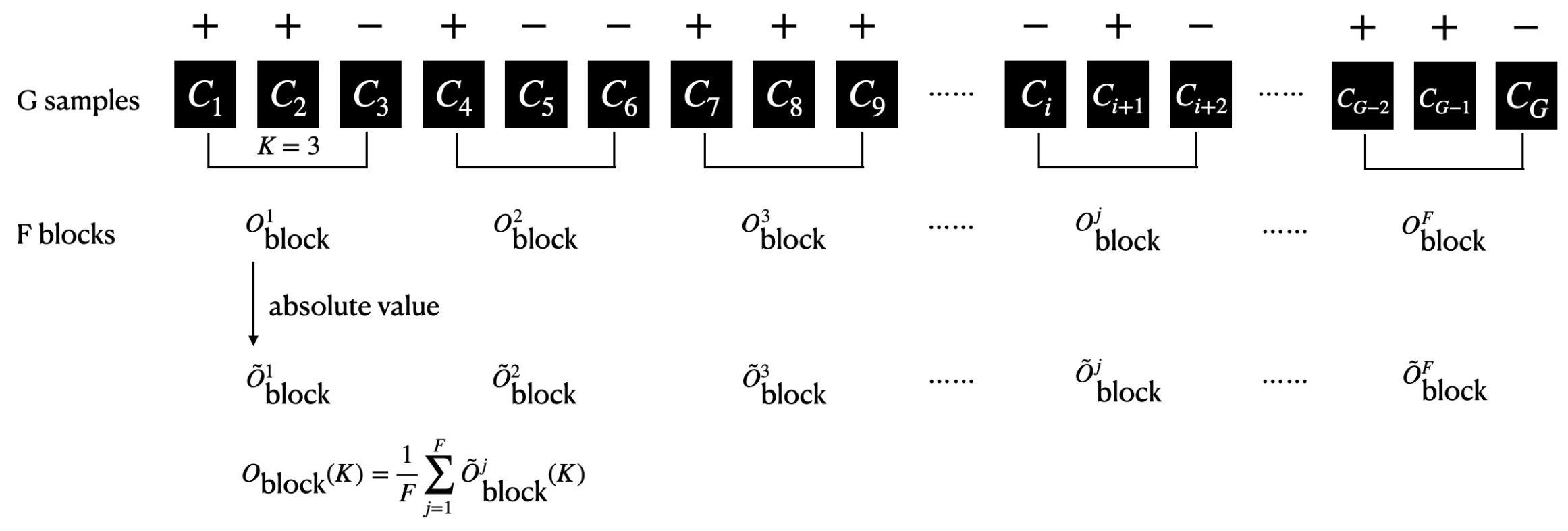}
\caption{\label{illublock}Schematic representation of obtaining $O_{\text{block}}(K)$ through the sign-blocking method. For the $G$ signed samples shown in the figure (where the sign factors are indicated above the black squares), we use $K=3$ as an illustrative example to demonstrate the statistical procedure for calculating the block estimator $O_{\text{block}}(K)$.}
\end{center}
\end{figure}

In this paper, we propose a sign-blocking method that partially captures the interference between positive and negative weight regions while circumventing the numerical instability of the sign problem. By introducing a block size $K$, we partition the total sample set into $\mathcal{F}$ blocks. Assuming $\langle O \rangle> 0$, we first compute the standard reweighted average for the $j$-th block:
\begin{equation}
{O}_{\text{block}}^j(K)=\frac{{\sum\limits^{(j)}}_i O_i S_i}{{\sum\limits^{(j)}}_i S_i}.
\end{equation}
Here, the summation ${\sum\limits^{(j)}}_i$ denotes the sum over all $K$ samples within the $j$-th block.
Inspired by the sign-ignoring method (Eq.~(\ref{k1})), which is free of the sign problem, we heuristically modify this expression as follows:
\begin{equation}
\tilde{O}_{\text{block}}^j(K) = \left| \frac{{\sum\limits^{(j)}}_i O_i S_i}{{\sum\limits^{(j)}}_i S_i} \right|.
\label{absoluteV}
\end{equation}
If there are $\mathcal{F}\equiv G/K$ blocks in total, we may explore the final estimator:
\begin{equation}
O_{\text{block}}(K) = \frac{1}{\mathcal{F}} \sum_{j=1}^{\mathcal{F}} \tilde{O}_{\text{block}}^j(K).
\label{OK}
\end{equation}
We note that $ {O} _{\text{ignore}}$ corresponds to the specific case where the block size is unity, \textit{i.e.}, $O_{\text{block}}(K=1)$. In Fig. \ref{illublock}, we present a schematic illustration of the procedure for obtaining the block estimator $O_{\text{block}}(K)$ from a set of $G$ signed samples. In this schematic, a block size of $K=3$ is chosen as an illustrative example.

For cases where $\langle O \rangle < 0$, the value of the observable can be inferred by simply applying a negative sign to the final result obtained from the sign-blocking procedure.

We now assume that the fermionic property $O_F(N)$ can be approximated as follows:
\begin{equation}
    O_F(N)\approx O_F(N,\alpha)=O_{\text{block}}(N,K=1)\pm\{O_{\text{block}}(N,K=1)-O_{\text{block}}(N,K=f(N))\},
    \label{correction_eq}
\end{equation}
where $N$ denotes the system size (\textit{e.g.}, $N = L_x \times L_y$ for a 2D Fermi-Hubbard model). The first term on the right-hand side represents the leading-order contribution obtained by neglecting the sign, while the second term accounts for the higher-order corrections that arise from the correlation between energy and sign. Since the absolute value operation is applied during the simulation of $O_{\text{block}}(N,K)$, an ambiguity arises regarding the sign of the aforementioned higher-order correction term.
By defining the function $f(N)$ as an \textit{ansatz} with tunable parameters, we can determine the optimal value of these tunable parameters for a small system through comparison with exact results. In the process of comparing our result for small system against the exact value, we can also unambiguously determine whether the higher-order correction term should take a positive or negative sign. Using the fixed parameters in $f(N)$, Eq. (\ref{correction_eq}) is then applied to larger-scale systems. Compared to the method outlined in Eq. (\ref{simple}), Eq. (\ref{correction_eq}) represents an upgraded approximation scheme. 

Since the fermion sign problem vanishes at vacuum (\textit{i.e.}, $N=0$), we may assume a general form for $f(N)$ given by:
\begin{equation}
    f(N)=\alpha N+\sum_{j>1}\beta_jN^j+1.
    \label{f(N)}
\end{equation}
In this case, at $N=0$, the second term on the right-hand side of Eq. (\ref{correction_eq}) formally cancels out. While the polynomial expansion employed above represents the most straightforward approach, we cannot rule out the possibility that more suitable function may be identified in future studies across diverse fermionic systems.

For the fermionic systems under investigation, to ensure that the sign-blocking method operates in its most straightforward manner, it is desirable for Eq. (\ref{f(N)}) to provide accurate predictions by retaining only the linear term. Consequently, we adopt the following approximation in this work:
\begin{equation}
f(N)=\alpha N+1.
\label{f(N)app}
\end{equation}
Remarkably, we find that for the 2D Fermi-Hubbard model studied here, this minimalist form yields remarkably reliable energy results. We do not, however, claim that this simplified expression is universally applicable to all fermionic systems.

We now summarize the procedure for the sign-blocking method analyzed above:
\begin{enumerate}
    \item For a small system $N_s$, we obtain a series of $O_{\text{block}}(N_s, K)$ for different $K$ through simulation.
    \item The parameter $\alpha$ is determined by calibrating the results of Eq. (\ref{correction_eq}) against the exact value $O_F(N_s)$, or high-accuracy benchmarks obtained from state-of-the-art methods, for small systems.
    \item This $\alpha$ is then applied to larger systems, where the simulated $O_F(N, \alpha)$ is taken as the approximation of the true fermionic properties.
\end{enumerate}
Consequently, this approach represents an inherent enhancement of Eq. (\ref{simple}), as the sign-blocking mechanism allows us to consider the interference among different signs.

It is worth noting that the proposed sign-blocking method shares an identical simulation procedure with the standard reweighting method; the distinction lies solely in the post-processing of the generated signed samples. Consequently, the method is highly efficient for high-performance computing, as it can be applied to existing state-of-the-art simulation method to get the signed samples.

To ensure the numerical stability of $O_{\text{block}}(K)$, we restrict the block size $K$ to odd integers in our practical applications, as even values of $K$ may lead to a vanishing denominator. In cases where the sign problem is most severe, the average sign within a block of size $K$ scales as $S \sim 1/K$. While a large $K$ still requires a substantial number of samples, this algebraic scaling represents a fundamental enhancement in computational cost, compared to the exponential decay of the average sign encountered in the conventional reweighting method.

\section{Application of the sign-blocking method to the 2D Fermi-Hubbard model}
\label{Appl}

\subsection{Background}

The numerical study of the Fermi-Hubbard model \cite{imada1998metal,arovas2022hubbard,qin2022hubbard} stands as a cornerstone in condensed matter physics, offering profound insights into high-temperature superconductivity and strongly correlated electronic systems. However, the notorious fermion sign problem \cite{dornheim2019fermion} remains a primary obstacle for stochastic simulation methods like  DQMC\cite{blankenbecler1981monte,hirsch1985two,lin1987two,hirsch1988pairing,white1989numerical,sorella1989novel,loh1990sign,santos2003introduction,huang2017numerical,assaad2025alf,song2025extended}. This problem manifests as an exponential decay of the average sign with increasing system size and inverse temperature, leading to an exponential bottleneck where statistical noise overwhelms the physical signal. Over the past decade, substantial advancements have been made in the numerical simulation of the Fermi-Hubbard model. These include systematic benchmarking efforts \cite{leblanc2015solutions}, the development of DMRG and fermionic PEPS \cite{verstraete2008matrix,Liu_PRB,WY_Liu} for two-dimensional systems \cite{liu2025accurate}, and the application of constrained-path AFQMC (CP-AFQMC) to access lower temperatures in doped regimes \cite{zheng2017stripe}. Furthermore, significant progress has been made through DMET \cite{knizia2012density} and the novel fictitious particle Hubbard model \cite{fan2025quantum} by extending isothermal $\xi$-extrapolation for continuous systems \cite{xiong2022thermodynamic,dornheim2023fermionic,dornheim2025unraveling}.

Traditional approaches to mitigate this issue for Fermi-Hubbard model, such as the fixed-node (FN) \cite{fahy1990positive} or constrained-path (CP) methods \cite{zhang1995constrained,zhang1999finite,zhang2003quantum}, often rely on a \textit{priori} assumptions about the nodal structure of the many-body wavefunction. While effective, these methods can introduce variational bias. Consequently, it is essential to develop approaches that do not rely on a trial wavefunction. Despite substantial advancements in simulation methods for the Fermi-Hubbard model over the past decade, several prototypical cases remain where different methodologies yield significantly inconsistent results, necessitating the development of alternative approaches for adjudication. For instance, in the square lattice at $U=8$ and $n=0.875$, the ground-state energies from FN, CP-AFQMC, and DMRG display discernible differences in the thermodynamic limit, with approximately 1\% variance between competing benchmarks \cite{leblanc2015solutions}. We will demonstrate that the sign-blocked estimator effectively validates energy benchmarks for the Fermi-Hubbard model, even in regimes where traditional methods exhibit obvious variance.

\subsection{The Fermi-Hubbard model and the operational procedure of the sign-blocking method}

We define the single-band Hamiltonian for the Fermi-Hubbard model as:
\begin{equation}
\hat H = {- t\sum_{\langle i,j\rangle,\sigma} \left(\hat c^{\dagger}_{i\sigma}\hat c_{j\sigma}+\text{H.c.}\right)} + {U\sum_i (\hat n_{i\uparrow}-\frac{1}{2}) (\hat n_{i\downarrow}-\frac{1}{2})}- \mu\sum_{i,\sigma}\hat n_{i\sigma}.
\end{equation}
The parameters $t$, $U$, and $\mu$ denote the nearest-neighbor hopping amplitude, the on-site interaction strength, and the chemical potential, respectively. We restrict the kinetic term to nearest-neighbor hopping, denoted by $\langle i, j \rangle$.

Under the AFQMC formalism, after applying the Hubbard-Stratonovich (HS) transformation and exactly tracing out the fermionic degrees of freedom for an arbitrary auxiliary field configuration $\{s\}$, the partition function of the Fermi-Hubbard model becomes:
\begin{equation}
    Z = \sum_{\{s\}} \det M_\uparrow(\{s\}) \det M_\downarrow(\{s\})=\sum_{\{s\}} W(\{s\})=\sum_{\{s\}} S(\{s\})\left|W(\{s\})\right|.
\end{equation}
Here, $\{s\}$ represents the auxiliary HS field configuration. For a 2D $L_x \times L_y$ lattice with $N = L_xL_y$ sites, if we employ the discrete HS transformation to the interaction term, we introduce a set of Ising-like auxiliary variables $s_{i,l} = \pm 1$ at each lattice site $i \in \{1, \dots, N\}$ and each imaginary-time slice $l \in \{1, \dots, M\}$ (where $M = \beta / \Delta\tau$). Consequently, the complete configuration $\{s\}$ is a space-time matrix of size $N \times M$, and the summation $\sum_{\{s\}}$ runs over all $2^{N \times M}$ possible configurations.

DQMC is a specific implementation of the AFQMC method as applied to the Fermi-Hubbard model. A significant advantage of DQMC is that it is free from the fermion sign problem \cite{hirsch1985two} at half-filling ($\mu = 0$). However, away from half-filling ($\mu \neq 0$), the sign problem re-emerges because the weights $W(\{s\})$ are no longer guaranteed to be positive-definite. Consequently, these weights can no longer be interpreted as classical probability distributions.

We now consider the practical implementation of the sign-blocking method for 2D Fermi-Hubbard model under periodic boundary conditions in both spatial dimensions. For a given set of parameters $\{U, \beta, n\}$—where $n$ denotes the average particle filling per site, numerical simulations to get the samples are performed using the standard DQMC method.

For a $L_x \times L_y$ lattice, let $E_F(N)$ denote the total energy of the fermionic system. We define $E_{\text{block}}(N, K=1)$ as the energy obtained by neglecting the sign. Based on the sign-blocking method, the fermionic energy is estimated as:
\begin{equation}
    E_F(N)\approx E_{\text{block}}(N,K=1)+\Delta E_{\text{block}}(N,K_\alpha),
    \label{correction_energy}
\end{equation}
where
\begin{equation}
    \Delta E_{\text{block}}(N,K_\alpha)=\pm\{E_{\text{block}}(N,K=1)-E_{\text{block}}(N,K_\alpha=f(N))\}.
\label{blockpn}
\end{equation}
In Eq. (\ref{correction_energy}), the term in the curly brackets represents the sign-driven correction to the $K=1$ case.

Based on the general consideration in Sec. \ref{SigIntr}, we assume
\begin{equation}
f(N)=\alpha L_xL_y+1.
\label{fxy}
\end{equation}
While the coefficient $\alpha$ is not known a \textit{priori}, our strategy involves determining the optimal $\alpha$ by benchmarking against exact result \cite{dagotto1992static} for small system. This calibrated $\alpha$ is then employed to predict the properties of larger systems. We note that this scaling \textit{ansatz} is specifically designed for homogeneous systems.

\section{Results}
\label{Results}

\subsection{Square lattice}

We focus on a representative case with parameters $U=8$ and $n=0.875$ (corresponding to $1/8$ hole doping) on a square lattice of width $L$. We set $\beta=16$ to approximate the system as the ground state, whereas finite-temperature cases are beyond the scope of this work.

\begin{figure}[htbp]
\includegraphics[scale=0.5]{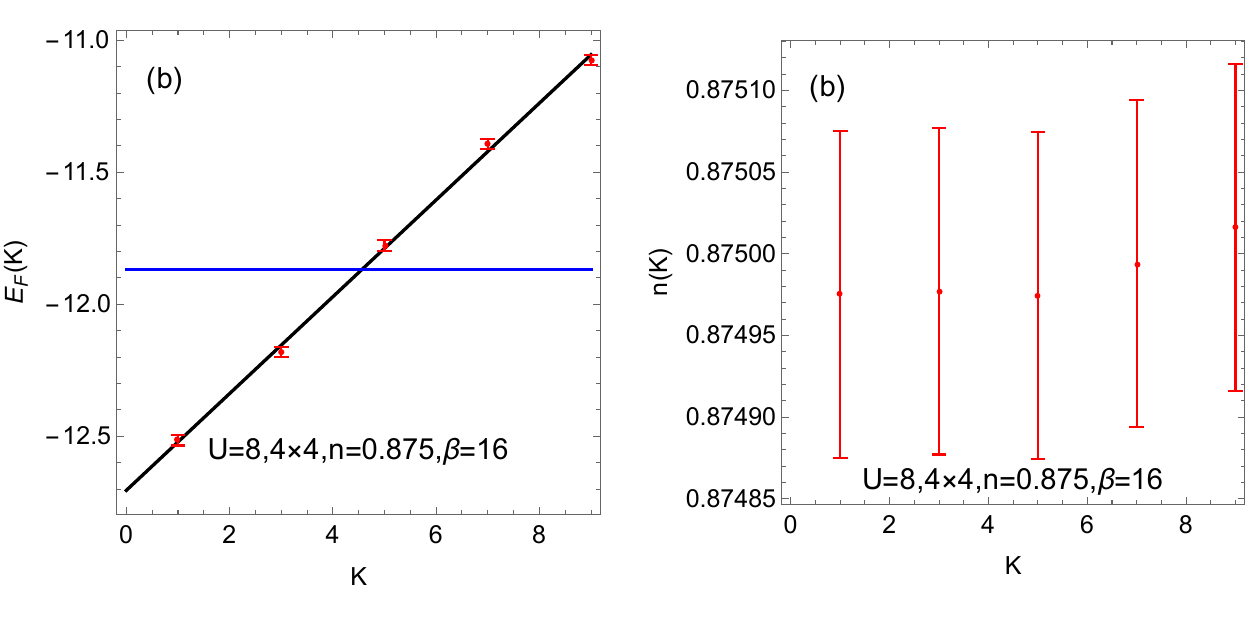}
\begin{center}
\caption{\label{L44U8}For $U=8,4\times 4,n=0.875,\beta=16$: (a) Red points with error bars represent the energy $E_F(K)$ for different blocking sizes, with the black line indicating a linear fit. The horizontal blue line denotes the result of ED \cite{dagotto1992static}. (b) Red points with error bars represent the average occupancy $n(K)$ for different blocking sizes.}
\end{center}
\end{figure}

In Fig. \ref{L44U8}(a), for $n=0.875$, the red points with error bars represent the energy $E_F(K)=2E_{\text{block}}(K=1)-E_{\text{block}}(K)$ for various block sizes $K$, with the blue horizontal line indicating the exact result obtained from ED \cite{dagotto1992static}. In Fig. \ref{L44U8}(b), we present the average particle filling per site, $n(K)$, as a function of $K$. We observe that the dependence of $n(K)$ on the block size $K$ is essentially negligible. This suggests that the target filling $n$ can be determined by tuning the chemical potential $\mu$, without requiring additional calibration for the average occupancy, in contrast to the energy, within the sign-blocking framework.

To ensure that $\alpha$ remains solvable in the $K > 0$ regime, the positive branch of the correction term in Eq.~(\ref{blockpn}) must be chosen. In Fig. \ref{L44U8}(a), the black line represents a linear fit to the energy data $E_F(K)$, from which we extract the optimal block size $K_c = 4.56$ (where $E_F(K_c)$ matches the ED benchmark \cite{dagotto1992static}). 
The observed strong linear relationship between $E_F(K)$ and $K$ enables the data points in Fig. \ref{L44U8}(a) to be characterized by a single fitting parameter. This observation provides empirical justification for the one-parameter functional form assumed in Eq. (\ref{fxy}).
Substituting this value into our scaling \textit{ansatz}, $f(L) = \alpha L^2 + 1$, and setting $f(L=4) = K_c = 4.56$, we determine the calibration coefficient to be $\alpha = 0.2225$.

\begin{figure}[htbp]
\begin{center}
\includegraphics[scale=0.6]{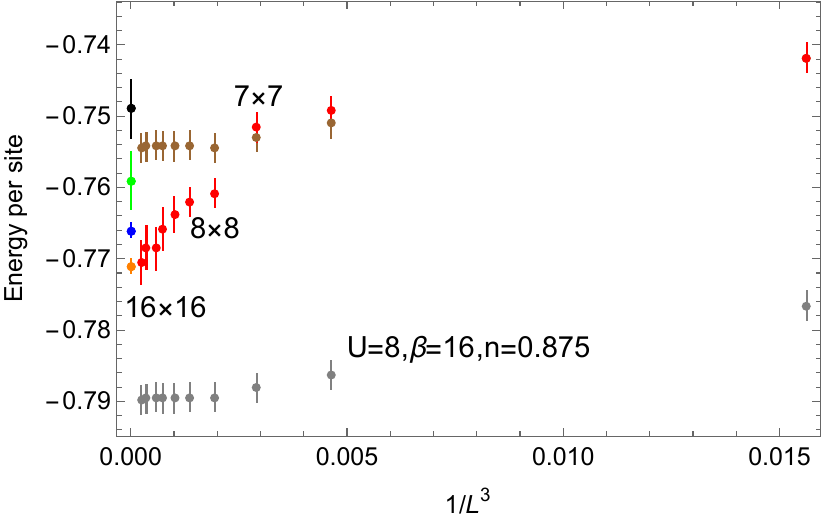}
\caption{\label{U8differentL}For square lattices with $U=8,\beta=16,n=0.875$, the red points with error bars represent the energy per site inferred by the sign-blocking method, while the grey points denote the energy per site obtained by ignoring the signs. The three annotated red points represent the $7\times 7$, $8\times 8$, and $16\times 16$ lattice systems, with the latter corresponding to the leftmost red point. The black, green, blue and orange points correspond to the ground-state energy benchmarks in the thermodynamic limit obtained via FN, DMRG, CP-AFQMC, and DMET, respectively \cite{leblanc2015solutions,zheng2017stripe}. The brown points represent the correction scheme (Eq. (\ref{simple})) derived from comparing the sign-ignored results with the exact results on a $4\times 4$ lattice.}
\end{center}
\end{figure}

In Fig. \ref{U8differentL}, we apply the sign-blocking method to larger square lattices. The red points with error bars represent the energy per site obtained via the sign-blocking method, while the grey points with error bars denote the results obtained by neglecting the sign ($K=1$). A distinct discrepancy is observed between the red and grey points. While the grey points reach a plateau for $1/L^3 \leq 3 \times 10^{-3}$, the red points continue to decrease monotonically. This indicates that the fermionic energy exhibits a more pronounced finite-size dependence compared to the approximate energy obtained by ignoring the sign. For comparison, we have included benchmark  ground-state energies in the thermodynamic limit from other numerical methods: black point for the FN, green point for DMRG, blue point for CP-AFQMC \cite{leblanc2015solutions}, and orange point for DMET \cite{zheng2017stripe}. Since the iPEPS results are closely aligned with those of CP-AFQMC, they are omitted from the figure for clarity; specific values can be found in Ref. \cite{zheng2017stripe}. 
Fig. \ref{U8differentL} clearly illustrates the formidable challenge posed by the Fermi-Hubbard model, as discernible discrepancies persist among the ground-state energies obtained by various state-of-the-art numerical methods.

The results obtained by neglecting the sign (grey points) deviate significantly from the expected physical values. In contrast, the sign-blocking method yields remarkably accurate results despite its procedural simplicity. Specifically, our results demonstrate a high degree of consistency with the DMET benchmarks \cite{zheng2017stripe}, outperforming the FN and DMRG estimates in this specific parameter regime. The energy obtained via the sign-blocking method is slightly lower than the CP-AFQMC benchmark; this is consistent with expectations, as CP-AFQMC provides a variational upper bound to the exact energy.

The most striking feature in Fig. \ref{U8differentL} is the abrupt decrease in energy per site observed when transitioning from the $7\times 7$ to the $8\times 8$ lattice. The sudden reduction in energy per site at the $8 \times 8$ lattice suggests a potential emergence of spatial correlations, such as stripe order. Although the energy per site is insufficient to confirm a stripe phase, the sudden drop observed at $8\times 8$ strongly supports the formation of a new ordered state. Notably, the sign-blocking approach has the advantage of capturing such phenomena without predefined symmetry breaking or assuming a stripe pattern beforehand.

For comparison, Fig. \ref{U8differentL} displays brown points with error bars representing the energy corrections for larger lattice systems by Eq. (\ref{simple}). These corrections are derived by calculating the relative deviation between the exact energy and the sign-ignored energy at the $4\times 4$ lattices, a strategy justified by the homogeneity of the periodic Fermi-Hubbard model. We observe that simply extrapolating these large-lattice results to the thermodynamic limit yields values that even outperform the FN method. 
Nevertheless, it is evident that the sign-blocking method exhibits significantly superior performance, compared with this simple correction scheme.

\begin{figure}[htbp]
\begin{center}
\includegraphics[scale=0.7]{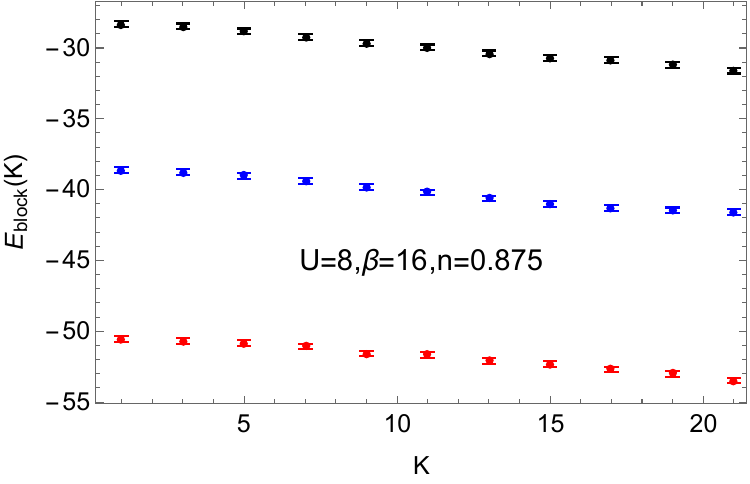}
\caption{\label{EKL8}For the parameters $U=8, \beta=16$, and $n=0.875$, we present the energy $E_{\text{block}}(K)$ for three different lattice sizes: $6 \times 6$ (black points), $7 \times 7$ (blue points), and $8 \times 8$ (red points). We observe a simple monotonic decrease in the energy $E_{\text{block}}(K)$ as the block size $K$ increases.}
\end{center}
\end{figure}

In Fig. \ref{EKL8}, we present the energy $E_{\text{block}}(K)$ for $6\times 6$ (black points), $7\times 7$ (blue points) and $8 \times 8$ (red points) lattice system. We observe that $E_{\text{block}}(K)$ exhibits a consistent and smooth dependence on $K$. This systematic behavior provides intuitive support for the reliability of the sign-blocking method, suggesting that $E_{\text{block}}(K)$ is successfully capturing the underlying physical correlations as the block size increases.

\subsection{Rectangular lattice}

We now perform numerical experiments to test whether the sign-blocking method, supplemented by Eq. (\ref{fxy}), can be extended to rectangular lattices. We specifically examine the $L_x \times L_y$ geometry with $L_x = 4$ at $U=8$ and $\beta=16$. This case is particularly interesting as it falls within the regime where DMRG is highly effective and yields results in excellent agreement with CP-AFQMC \cite{zheng2017stripe}. Furthermore, this system is known to predict the existence of a period-8 stripe phase. 

\begin{figure}[htbp]
\begin{center}
\includegraphics[scale=0.6]{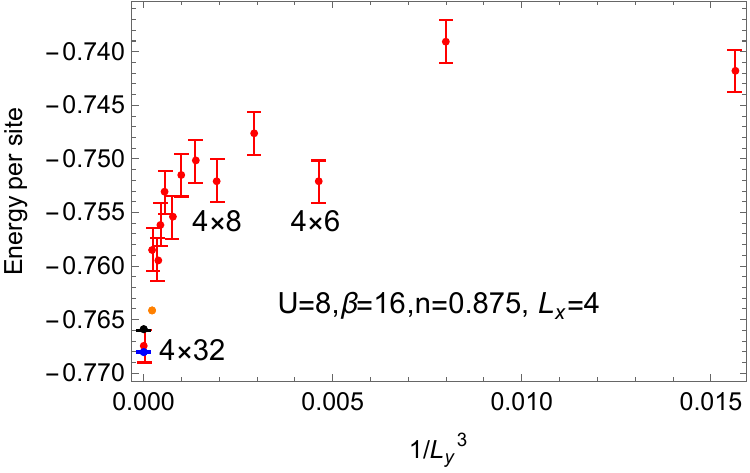}
\caption{\label{L4width}For the Fermi-Hubbard model with a width of $L_x=4$ and $U=8,\beta=16,n=0.875$, the red points represent the energy per site obtained by the sign-blocking method. The black and blue points denote the ground-state energies in the thermodynamic limit obtained by DMRG and CP-AFQMC, respectively \cite{zheng2017stripe}. The orange point indicates the energy per site for the $4 \times 16$ lattice \cite{liang2025investigating} calculated using the tensor-backflow method \cite{zhou2024solving}.}
\end{center}
\end{figure}

In Fig. \ref{L4width}, for the parameters $U=8, \beta=16, n=0.875, L_x=4$, we present the energy per site for different $L_y$, where the red points with error bars represent the results from the sign-blocking method. The black and blue points denote the ground-state energies in the thermodynamic limit obtained by DMRG and CP-AFQMC, respectively \cite{zheng2017stripe}. The orange point indicates the energy per site for the $4 \times 16$ lattice \cite{liang2025investigating} calculated using the tensor-backflow method \cite{zhou2024solving}. No error bar was provided for the orange point in Ref. \cite{liang2025investigating}. In this figure, we observe a dip in the energy per site for the $4\times8$ and $4\times6$ lattices. Our simulations extend up to a maximum size of $4\times32$, where the result is highly consistent with CP-AFQMC within the range of fluctuations. In contrast to the square lattice analyzed previously, the DMRG results for the long-strip geometry exhibit much better agreement with the sign-blocking method. This is consistent with the prevailing consensus regarding the applicability and strengths of DMRG.

\begin{figure}[htbp]
\begin{center}
\includegraphics[scale=0.4]{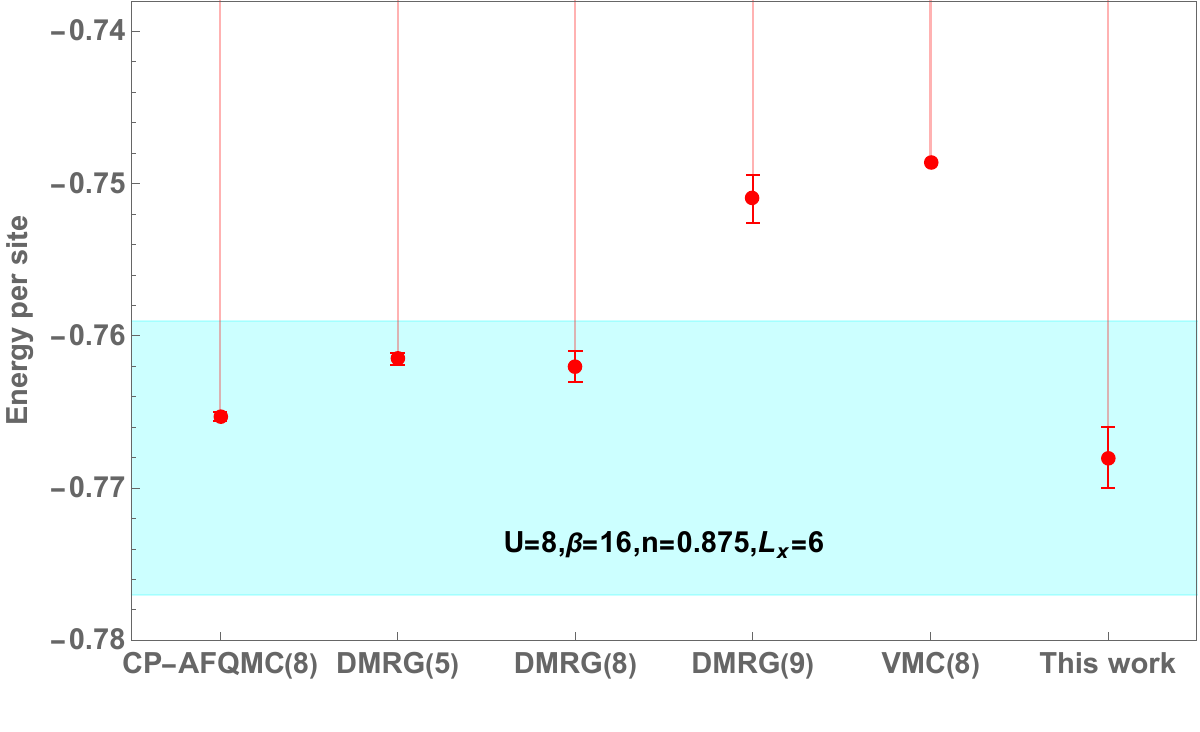}
\caption{\label{L6width}For the Fermi-Hubbard model with a width of $L_x=6$ and $U=8,\beta=16,n=0.875$, shown are the energy per site for different methods.}
\end{center}
\end{figure}

We now turn our attention to the case of $L_x=6, U=8, n=0.875$, and $\beta=16$. In Fig. \ref{L6width}, the label “VMC” represents the ground-state energy per site for a $6 \times 16$ lattice obtained via variational Monte Carlo \cite{marino2022stripes}, “This work” is the sign-blocking method for $6\times 32$, whereas all other data correspond to results in the thermodynamic limit. The values in parentheses on the $x$-axis denote the predefined periods of the stripe phase used in those specific simulations; these CP-AFQMC and DMRG benchmarks are taken from Ref. \cite{zheng2017stripe}. In the present work, no period is labeled as the sign-blocking method does not require an a \textit{priori} assumption regarding the stripe phase. 

Fig. \ref{L6width} underscores the significant value of the sign-blocking method. Without requiring any a \textit{priori} assumptions regarding the properties of the Fermi-Hubbard model, this approach effectively rules out the DMRG results based on a 9-period stripe phase, while remaining consistent with the more plausible 5-period and 8-period results. Furthermore, we note that the results obtained via the sign-blocking method are more reasonable than the result from recently developed variational Monte Carlo techniques \cite{marino2022stripes}. Given its predictive power and independence from model-specific assumptions, we believe this method can offer meaningful contributions to the systematic analysis and benchmarking of Fermi-Hubbard models in the future.

\section{Strategy for Implementing the Sign-Blocking Method in AFQMC}
\label{Strategy}

The practical implementation of the proposed sign-blocking method begins with generating a large ensemble of signed stochastic samples using quantum Monte Carlo (QMC). It is important to note that these signed samples are intrinsically dependent on the specific QMC algorithm employed. While the preceding results were obtained within the auxiliary-field DQMC framework for the Fermi–Hubbard model, we now turn to an alternative QMC approach that produces signed samples via the fermionic propagator method.

Within path-integral approach, we employ the fermionic propagator \cite{Raedt1981,Dornheim_2015,Chin2024,xiong2025pseudo} to bridge consecutive time slices. While this representation avoids the sign problem in 1D \cite{Chin2024}, it does not offer the same protection for the 2D Fermi-Hubbard model. Notably, the fermion sign problem re-emerges in the 2D case, regardless of whether the system is at half-filling.

\begin{figure}[htbp]
\begin{center}
\includegraphics[scale=0.7]{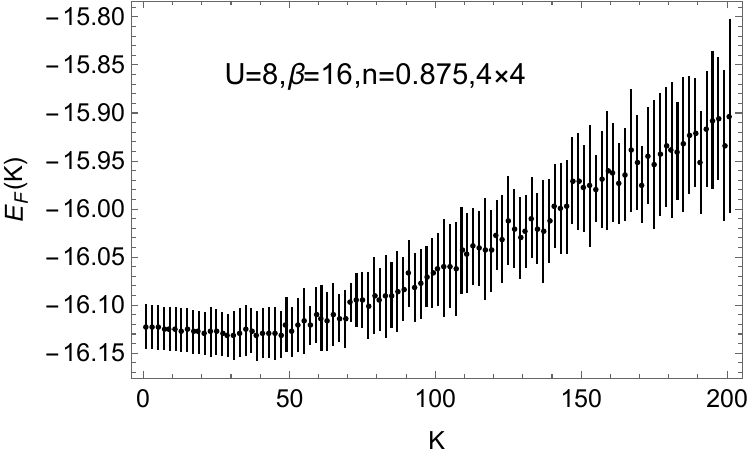}
\caption{\label{Fpropagator}For a $4\times4$ lattice with $U=8, \beta=16$, and $n=0.875$, the black points with error bars represent $E_F(K)$ obtained via the sign-blocking method following fermionic propagator PIMC sampling. For these parameters, the exact ground-state energy \cite{dagotto1992static} of the fermionic system is $E_g = -11.86884$.}
\end{center}
\end{figure}

To further delineate the applicability of the sign-blocking method, we generate signed samples for the 2D Fermi-Hubbard model using the fermionic propagator method instead of DQMC. For the case of $L_x=L_y=4, n=0.875, \beta=16$, Fig. \ref{Fpropagator} illustrates the relationship between the energy $E_F(K)$ and the block size $K$, while the exact ground-state energy \cite{dagotto1992static} of the fermionic system is $E_g = -11.86884$. Unfortunately, we observe only a weak dependence of $E_F(K)$ on $K$; moreover, the result at $K=1$ exhibits a significant discrepancy from the exact ground-state energy. Within the observed $K < 200$, no intersection is found between $E_F(K)$ and the exact value $E_g$; furthermore, the statistical fluctuations intensify as $K$ increases.
This suggests that when paired with the fermionic propagator method, the sign-blocking approach is unable to effectively infer the thermodynamic properties of the 2D Fermi-Hubbard model. However, this does not undermine the general utility of the method, as our previous results demonstrate that the sign-blocking method yields highly reliable energies when the signed samples are obtained via DQMC. The comparison between DQMC and the fermionic propagator method indicates that, when extending the sign-blocking method to other fermionic systems in the future, the AFQMC framework should be prioritized as the sampling engine for generating signed configurations. 

The distinction between DQMC and the fermionic propagator method can be elucidated through a thought experiment. In DQMC, the classical auxiliary HS field configuration, $\{s\}$, is coupled to the fermionic system. Once a specific configuration $\{s\}$ is fixed, the fermions and the auxiliary field can be treated as a unified composite system. 
Since the fermionic degrees of freedom have been exactly traced out, we can employ Monte Carlo sampling to “measure” the energy $E(\{s\})$ and the corresponding sign factor $S(\{s\})$ for the integrated system at each specific configuration. The configuration $\{s\}$ can be interpreted as a fictitious measurement apparatus probing both the energy and the phase (\textit{i.e.}, the sign factor).
To rigorously determine the total energy of the fermionic system, an analysis across the entire ensemble of configurations $\{s\}$ is required. At half-filling, DQMC cleverly ensures that the interference between different fermionic states results in a sign factor of unity for all configurations $\{s\}$. Unfortunately, away from half-filling, the sign factors “measured” in DQMC can be either positive or negative.
The sign-blocking method therefore operates directly on these “measured” results, namely the ensemble of signed samples obtained from the composite fermion–HS system.

\begin{figure}[htbp]
\begin{center}
\includegraphics[scale=0.29]{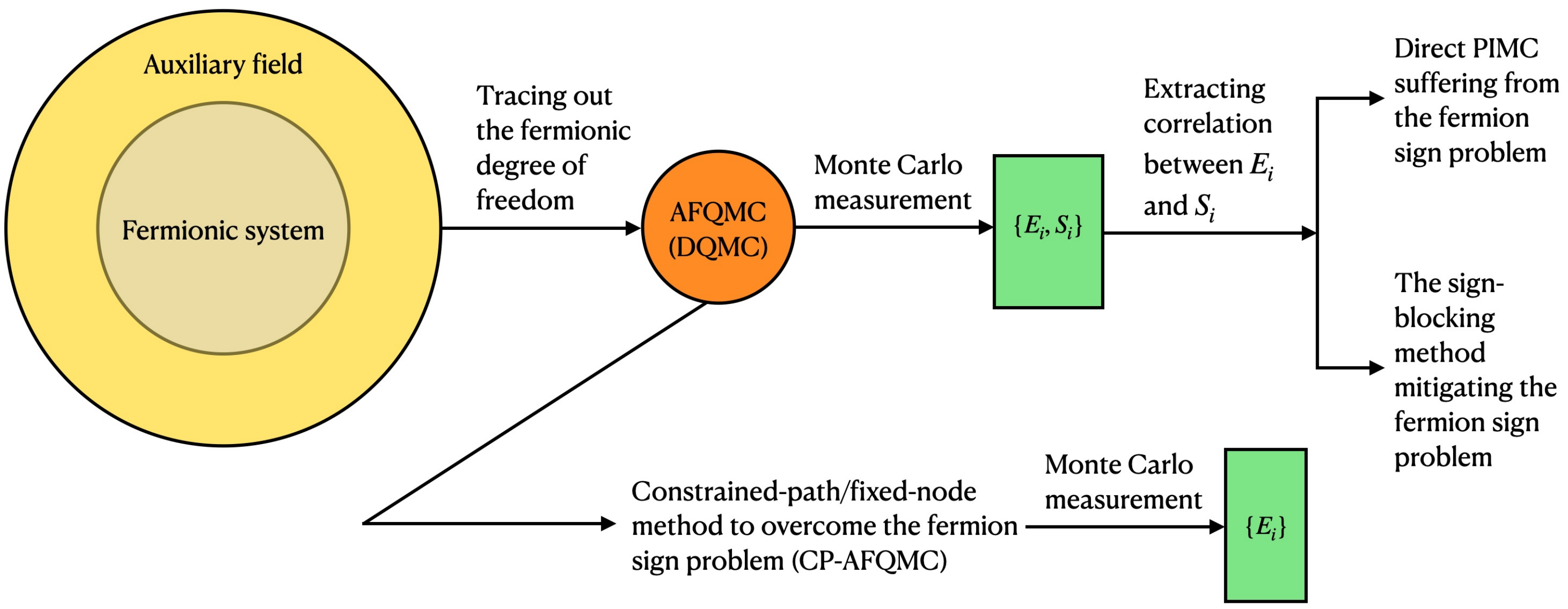}
\caption{\label{AFQMC} A general strategy for the implementation of the sign-blocking method. Both direct PIMC and the sign-blocking method are capable of extracting the correlation between the energy $E_i$ and the sign factor $S_i$. However, while direct PIMC is severely hindered by the fermion sign problem, the sign-blocking method significantly mitigates this issue.
For comparison, the schematic also illustrates CP-AFQMC, which employs path constraints to circumvent the sign problem.
}
\end{center}
\end{figure}

By contrast, in the fermionic propagator method, the energy and sign factor associated with a specific configuration in the path-integral space cannot be regarded as a Monte Carlo “measurement” of the fermionic system’s statistical average, because the fermionic degrees of freedom, namely the spatial coordinates at each imaginary-time slice, have not been traced out.
This stands in stark contrast to the DQMC approach, where the auxiliary field coupling allows for a quasi-physical interpretation of each sample. This structural difference is likely the root cause of the disparate performance of the sign-blocking method across these two engines. While it is premature to conclude that AFQMC is strictly essential for the sign-blocking method, these heuristic insights suggest that future applications should prefer AFQMC-based sampling to ensure compatibility with the underlying energy-sign correlation, as the fermionic degrees of freedom are traced out in this framework. Guided by the preceding intuitive analysis, we outline in Fig. \ref{AFQMC} a general strategy for implementing the sign-blocking method across various fermionic systems in future studies. For comparison, this figure also illustrates the conceptual approaches of direct PIMC and CP-AFQMC. The following section (Sec. \ref{correlation}) further elucidates the physical significance of the correlation extraction shown in the figure.

\section{The Sign-Blocking Method as a Correlation Extraction Technique}
\label{correlation}

If we examine only the distribution of sign factors, it appears to be purely stochastic noise. If we only consider the statistical distribution and the average of the energy data, we effectively obtain the energy associated with the sign-ignored result; however, this result is insufficient to capture the underlying physics of the fermionic system.
Fortunately, an intrinsic correlation exists between the sign and energy distributions, as each configuration yields a unique energy and a corresponding sign factor. It is precisely this correlation that enables us, in principle, to retrieve exact fermionic observables via direct PIMC for large-scale systems, provided that computational resources are sufficiently vast.

In the sign-blocking method, we process a large volume of classical signals $\{E_j, S_j\}$ from Monte Carlo measurements using data blocking to recover intrinsic correlations between $E_j$ and $S_j$, and extract the quantum properties of the fermionic system. This concept, in which hidden quantum correlations are extracted from classical data, is well-established in other fields. For instance, in ultracold Fermi gas experiments \cite{deJongh2025,Yao2025}, the density distributions obtained from repeated shot-by-shot measurements are post-processed to infer the pair correlation function $g_2(r)$. The quantum statistical nature of fermions is clearly captured by the vanishing of $g_2(r)$ at short distances ($r \rightarrow 0$). For ultracold Fermi gases released from optical lattices, the quantum statistical properties can be accurately retrieved by extracting the correlation information from noise in the density distribution \cite{folling2005spatial}. Drawing on this analogy, the sign-blocking method represents a novel paradigm in the study of the fermion sign problem, utilizing correlation analysis of classical Monte Carlo signals to extract the fundamental properties of fermionic systems. In Fig. \ref{AFQMC}, we present a comparison between direct PIMC and the sign-blocking method, highlighting their shared characteristics as well as their differences.

The sign-blocking method specifically attempts to unveil the correlation between energy and sign factors through data blocking. In the context of the Fermi-Hubbard model, we observe that the sign-blocking method effectively reveals this correlation within the DQMC framework. Conversely, within the fermionic propagator framework, the method fails to expose this underlying connection. Consequently, the key to the successful application of the sign-blocking method in future research lies in developing techniques to effectively manifest the inherent correlation between energy and sign factors. Looking ahead, it is imperatively important to explore more diverse and sophisticated methodologies to extract these correlations beyond direct PIMC and the sign-blocking method, as they represent a largely untapped resource for overcoming the fermion sign problem in complex quantum systems.

\section{Summary and Discussion}
\label{Summary}

In summary, we have introduced and demonstrated the sign-blocking method as a robust tool for mitigating the fermion sign problem in the 2D Fermi-Hubbard model. The core of the sign-blocking method lies in the observation that the sign-ignoring result is merely the $K=1$ limit of a more general block-dependent estimator $O_{\text{block}}(K)$. By increasing $K$ and applying a systematic scaling \textit{ansatz}, we can extract energies that traditional reweighting methods fail to reach due to the exponential decay of the signal-to-noise ratio. Within the sign-blocking framework, the average sign follows an algebraic $1/K$ dependence instead of decaying exponentially.
Our results for the representative 2D Fermi-Hubbard model with $1/8$ hole-doping case show that the sign-blocking method not only captures the correct energy trends but also matches the state-of-the-art, high-precision benchmark. In essence, the sign-blocking method can be viewed as a significantly enhanced reweighting framework designed to overcome the fermion sign problem by uncovering the intrinsic correlation between energy and sign factors. Unlike conventional reweighting schemes, it partially accounts for the interference of the weights between different signs. Given its conceptual simplicity, the sign-blocking method offers a promising new avenue for the numerical exploration of strongly correlated quantum matter.

In this work, we employ the grand canonical DQMC algorithm, which requires adjusting the chemical potential to achieve the desired average filling. Fortunately, in the cases considered here, the average occupancy is largely independent of the blocking size. A promising direction for future research would be the integration of the sign-blocking method with canonical DQMC \cite{shen2020finite,shen2023stable}. In such a framework, the particle number is fixed from the outset, potentially leading to more accurate and simple inferences of fermionic properties.

Extending beyond the square-lattice Hubbard model, this approach is readily applicable to a broad spectrum of challenging systems. These include continuous quantum systems, such as the uniform electron gas \cite{dornheim2018uniform,dornheim2025reweighting,svensson2025accelerated,dornheim2025taylor,dornheim2025fermionic,dornheim2023fermionic} and $^3\text{He}$ \cite{ceperley1992path,morresi2025normal}, as well as fermions on frustrated geometries like triangular or kagome lattices, where the fermion sign problem is notoriously severe. 
It is worth noting that for continuum systems, incorporating the sign-blocking method into frameworks based on AFQMC appears to be a promising and viable path.
Furthermore, the method's ability to extract ground-state properties without the need for a trial wavefunction suggests its potential as a benchmark tool for other fixed-node, constrained-path or variational algorithms across diverse many-body contexts.

\section*{Acknowledgments}
Y. Xiong gratefully acknowledges the support of the Hubei Provincial Young Top-Talent Program for this work. We acknowledge Dr. Tommaso Morresi for helpful discussions.

\bibliography{biblio}

\begin{thebibliography}{71}%
\makeatletter
\providecommand \@ifxundefined [1]{%
 \@ifx{#1\undefined}
}%
\providecommand \@ifnum [1]{%
 \ifnum #1\expandafter \@firstoftwo
 \else \expandafter \@secondoftwo
 \fi
}%
\providecommand \@ifx [1]{%
 \ifx #1\expandafter \@firstoftwo
 \else \expandafter \@secondoftwo
 \fi
}%
\providecommand \natexlab [1]{#1}%
\providecommand \enquote  [1]{``#1''}%
\providecommand \bibnamefont  [1]{#1}%
\providecommand \bibfnamefont [1]{#1}%
\providecommand \citenamefont [1]{#1}%
\providecommand \href@noop [0]{\@secondoftwo}%
\providecommand \href [0]{\begingroup \@sanitize@url \@href}%
\providecommand \@href[1]{\@@startlink{#1}\@@href}%
\providecommand \@@href[1]{\endgroup#1\@@endlink}%
\providecommand \@sanitize@url [0]{\catcode `\\12\catcode `\$12\catcode
  `\&12\catcode `\#12\catcode `\^12\catcode `\_12\catcode `\%12\relax}%
\providecommand \@@startlink[1]{}%
\providecommand \@@endlink[0]{}%
\providecommand \url  [0]{\begingroup\@sanitize@url \@url }%
\providecommand \@url [1]{\endgroup\@href {#1}{\urlprefix }}%
\providecommand \urlprefix  [0]{URL }%
\providecommand \Eprint [0]{\href }%
\providecommand \doibase [0]{https://doi.org/}%
\providecommand \selectlanguage [0]{\@gobble}%
\providecommand \bibinfo  [0]{\@secondoftwo}%
\providecommand \bibfield  [0]{\@secondoftwo}%
\providecommand \translation [1]{[#1]}%
\providecommand \BibitemOpen [0]{}%
\providecommand \bibitemStop [0]{}%
\providecommand \bibitemNoStop [0]{.\EOS\space}%
\providecommand \EOS [0]{\spacefactor3000\relax}%
\providecommand \BibitemShut  [1]{\csname bibitem#1\endcsname}%
\let\auto@bib@innerbib\@empty
\bibitem [{\citenamefont {Loh~Jr}\ \emph {et~al.}(1990)\citenamefont {Loh~Jr},
  \citenamefont {Gubernatis}, \citenamefont {Scalettar}, \citenamefont {White},
  \citenamefont {Scalapino},\ and\ \citenamefont {Sugar}}]{loh1990sign}%
  \BibitemOpen
  \bibfield  {author} {\bibinfo {author} {\bibfnamefont {E.}~\bibnamefont
  {Loh~Jr}}, \bibinfo {author} {\bibfnamefont {J.}~\bibnamefont {Gubernatis}},
  \bibinfo {author} {\bibfnamefont {R.}~\bibnamefont {Scalettar}}, \bibinfo
  {author} {\bibfnamefont {S.}~\bibnamefont {White}}, \bibinfo {author}
  {\bibfnamefont {D.}~\bibnamefont {Scalapino}},\ and\ \bibinfo {author}
  {\bibfnamefont {R.}~\bibnamefont {Sugar}},\ }\bibfield  {title} {\bibinfo
  {title} {Sign problem in the numerical simulation of many-electron systems},\
  }\href@noop {} {\bibfield  {journal} {\bibinfo  {journal} {Physical Review
  B}\ }\textbf {\bibinfo {volume} {41}},\ \bibinfo {pages} {9301} (\bibinfo
  {year} {1990})}\BibitemShut {NoStop}%
\bibitem [{\citenamefont {Dornheim}(2019)}]{dornheim2019fermion}%
  \BibitemOpen
  \bibfield  {author} {\bibinfo {author} {\bibfnamefont {T.}~\bibnamefont
  {Dornheim}},\ }\bibfield  {title} {\bibinfo {title} {Fermion sign problem in
  path integral monte carlo simulations: Quantum dots, ultracold atoms, and
  warm dense matter},\ }\href {https://doi.org/10.1103/PhysRevE.100.023307}
  {\bibfield  {journal} {\bibinfo  {journal} {Phys. Rev. E}\ }\textbf {\bibinfo
  {volume} {100}},\ \bibinfo {pages} {023307} (\bibinfo {year}
  {2019})}\BibitemShut {NoStop}%
\bibitem [{\citenamefont {Troyer}\ and\ \citenamefont
  {Wiese}(2005)}]{troyer2005computational}%
  \BibitemOpen
  \bibfield  {author} {\bibinfo {author} {\bibfnamefont {M.}~\bibnamefont
  {Troyer}}\ and\ \bibinfo {author} {\bibfnamefont {U.-J.}\ \bibnamefont
  {Wiese}},\ }\bibfield  {title} {\bibinfo {title} {Computational complexity
  and fundamental limitations to fermionic quantum monte carlo simulations},\
  }\href@noop {} {\bibfield  {journal} {\bibinfo  {journal} {Physical Review
  Letters}\ }\textbf {\bibinfo {volume} {94}},\ \bibinfo {pages} {170201}
  (\bibinfo {year} {2005})}\BibitemShut {NoStop}%
\bibitem [{\citenamefont {Alexandru}\ \emph {et~al.}(2022)\citenamefont
  {Alexandru}, \citenamefont {Ba{\c{s}}ar}, \citenamefont {Bedaque},\ and\
  \citenamefont {Warrington}}]{alexandru2022complex}%
  \BibitemOpen
  \bibfield  {author} {\bibinfo {author} {\bibfnamefont {A.}~\bibnamefont
  {Alexandru}}, \bibinfo {author} {\bibfnamefont {G.}~\bibnamefont
  {Ba{\c{s}}ar}}, \bibinfo {author} {\bibfnamefont {P.~F.}\ \bibnamefont
  {Bedaque}},\ and\ \bibinfo {author} {\bibfnamefont {N.~C.}\ \bibnamefont
  {Warrington}},\ }\bibfield  {title} {\bibinfo {title} {Complex paths around
  the sign problem},\ }\href@noop {} {\bibfield  {journal} {\bibinfo  {journal}
  {Reviews of Modern Physics}\ }\textbf {\bibinfo {volume} {94}},\ \bibinfo
  {pages} {015006} (\bibinfo {year} {2022})}\BibitemShut {NoStop}%
\bibitem [{\citenamefont {Dornheim}\ \emph {et~al.}(2018)\citenamefont
  {Dornheim}, \citenamefont {Groth},\ and\ \citenamefont
  {Bonitz}}]{dornheim2018uniform}%
  \BibitemOpen
  \bibfield  {author} {\bibinfo {author} {\bibfnamefont {T.}~\bibnamefont
  {Dornheim}}, \bibinfo {author} {\bibfnamefont {S.}~\bibnamefont {Groth}},\
  and\ \bibinfo {author} {\bibfnamefont {M.}~\bibnamefont {Bonitz}},\
  }\bibfield  {title} {\bibinfo {title} {The uniform electron gas at warm dense
  matter conditions},\ }\href@noop {} {\bibfield  {journal} {\bibinfo
  {journal} {Physics Reports}\ }\textbf {\bibinfo {volume} {744}},\ \bibinfo
  {pages} {1} (\bibinfo {year} {2018})}\BibitemShut {NoStop}%
\bibitem [{\citenamefont {He}\ \emph {et~al.}(2026)\citenamefont {He},
  \citenamefont {Zeng}, \citenamefont {Yang}, \citenamefont {Wang},
  \citenamefont {Ye},\ and\ \citenamefont {Li}}]{he2025revisiting}%
  \BibitemOpen
  \bibfield  {author} {\bibinfo {author} {\bibfnamefont {R.-C.}\ \bibnamefont
  {He}}, \bibinfo {author} {\bibfnamefont {J.-X.}\ \bibnamefont {Zeng}},
  \bibinfo {author} {\bibfnamefont {S.}~\bibnamefont {Yang}}, \bibinfo {author}
  {\bibfnamefont {C.}~\bibnamefont {Wang}}, \bibinfo {author} {\bibfnamefont
  {Q.-J.}\ \bibnamefont {Ye}},\ and\ \bibinfo {author} {\bibfnamefont {X.-Z.}\
  \bibnamefont {Li}},\ }\bibfield  {title} {\bibinfo {title} {Fermion sign
  problem and the structure of {Lee-Yang} zeros: The form of the partition
  function for indistinguishable particles and its zeros at 0 k},\ }\href
  {https://doi.org/10.1103/m1py-qtt5} {\bibfield  {journal} {\bibinfo
  {journal} {Physical Review E}\ }\textbf {\bibinfo {volume} {113}},\ \bibinfo
  {pages} {024115} (\bibinfo {year} {2026})}\BibitemShut {NoStop}%
\bibitem [{\citenamefont {Wu}\ and\ \citenamefont
  {Zhang}(2005)}]{wu2005sufficient}%
  \BibitemOpen
  \bibfield  {author} {\bibinfo {author} {\bibfnamefont {C.}~\bibnamefont
  {Wu}}\ and\ \bibinfo {author} {\bibfnamefont {S.-C.}\ \bibnamefont {Zhang}},\
  }\bibfield  {title} {\bibinfo {title} {Sufficient condition for absence of
  the sign problem in the fermionic quantum monte carlo algorithm},\
  }\href@noop {} {\bibfield  {journal} {\bibinfo  {journal} {Physical Review
  B}\ }\textbf {\bibinfo {volume} {71}},\ \bibinfo {pages} {155115} (\bibinfo
  {year} {2005})}\BibitemShut {NoStop}%
\bibitem [{\citenamefont {Prokof’ev}\ and\ \citenamefont
  {Svistunov}(2007)}]{prokof2007bold}%
  \BibitemOpen
  \bibfield  {author} {\bibinfo {author} {\bibfnamefont {N.}~\bibnamefont
  {Prokof’ev}}\ and\ \bibinfo {author} {\bibfnamefont {B.}~\bibnamefont
  {Svistunov}},\ }\bibfield  {title} {\bibinfo {title} {Bold diagrammatic monte
  carlo technique: When the sign problem is welcome},\ }\href@noop {}
  {\bibfield  {journal} {\bibinfo  {journal} {Physical Review Letters}\
  }\textbf {\bibinfo {volume} {99}},\ \bibinfo {pages} {250201} (\bibinfo
  {year} {2007})}\BibitemShut {NoStop}%
\bibitem [{\citenamefont {Booth}\ \emph {et~al.}(2009)\citenamefont {Booth},
  \citenamefont {Thom},\ and\ \citenamefont {Alavi}}]{booth2009fermion}%
  \BibitemOpen
  \bibfield  {author} {\bibinfo {author} {\bibfnamefont {G.~H.}\ \bibnamefont
  {Booth}}, \bibinfo {author} {\bibfnamefont {A.~J.}\ \bibnamefont {Thom}},\
  and\ \bibinfo {author} {\bibfnamefont {A.}~\bibnamefont {Alavi}},\ }\bibfield
   {title} {\bibinfo {title} {Fermion monte carlo without fixed nodes: A game
  of life, death, and annihilation in slater determinant space},\ }\href@noop
  {} {\bibfield  {journal} {\bibinfo  {journal} {The Journal of chemical
  physics}\ }\textbf {\bibinfo {volume} {131}} (\bibinfo {year}
  {2009})}\BibitemShut {NoStop}%
\bibitem [{\citenamefont {Schoof}\ \emph {et~al.}(2011)\citenamefont {Schoof},
  \citenamefont {Bonitz}, \citenamefont {Filinov}, \citenamefont {Hochstuhl},\
  and\ \citenamefont {Dufty}}]{schoof2011configuration}%
  \BibitemOpen
  \bibfield  {author} {\bibinfo {author} {\bibfnamefont {T.}~\bibnamefont
  {Schoof}}, \bibinfo {author} {\bibfnamefont {M.}~\bibnamefont {Bonitz}},
  \bibinfo {author} {\bibfnamefont {A.}~\bibnamefont {Filinov}}, \bibinfo
  {author} {\bibfnamefont {D.}~\bibnamefont {Hochstuhl}},\ and\ \bibinfo
  {author} {\bibfnamefont {J.}~\bibnamefont {Dufty}},\ }\bibfield  {title}
  {\bibinfo {title} {Configuration path integral monte carlo},\ }\href@noop {}
  {\bibfield  {journal} {\bibinfo  {journal} {Contributions to Plasma Physics}\
  }\textbf {\bibinfo {volume} {51}},\ \bibinfo {pages} {687} (\bibinfo {year}
  {2011})}\BibitemShut {NoStop}%
\bibitem [{\citenamefont {Brown}\ \emph {et~al.}(2013)\citenamefont {Brown},
  \citenamefont {Clark}, \citenamefont {DuBois},\ and\ \citenamefont
  {Ceperley}}]{brown2013path}%
  \BibitemOpen
  \bibfield  {author} {\bibinfo {author} {\bibfnamefont {E.~W.}\ \bibnamefont
  {Brown}}, \bibinfo {author} {\bibfnamefont {B.~K.}\ \bibnamefont {Clark}},
  \bibinfo {author} {\bibfnamefont {J.~L.}\ \bibnamefont {DuBois}},\ and\
  \bibinfo {author} {\bibfnamefont {D.~M.}\ \bibnamefont {Ceperley}},\
  }\bibfield  {title} {\bibinfo {title} {Path-integral monte carlo simulation
  of the warm dense homogeneous electron gas},\ }\href@noop {} {\bibfield
  {journal} {\bibinfo  {journal} {Physical Review Letters}\ }\textbf {\bibinfo
  {volume} {110}},\ \bibinfo {pages} {146405} (\bibinfo {year}
  {2013})}\BibitemShut {NoStop}%
\bibitem [{\citenamefont {Li}\ \emph {et~al.}(2016)\citenamefont {Li},
  \citenamefont {Jiang},\ and\ \citenamefont {Yao}}]{li2016majorana}%
  \BibitemOpen
  \bibfield  {author} {\bibinfo {author} {\bibfnamefont {Z.-X.}\ \bibnamefont
  {Li}}, \bibinfo {author} {\bibfnamefont {Y.-F.}\ \bibnamefont {Jiang}},\ and\
  \bibinfo {author} {\bibfnamefont {H.}~\bibnamefont {Yao}},\ }\bibfield
  {title} {\bibinfo {title} {Majorana-time-reversal symmetries: A fundamental
  principle for sign-problem-free quantum monte carlo simulations},\ }\href
  {https://doi.org/10.1103/PhysRevLett.117.267002} {\bibfield  {journal}
  {\bibinfo  {journal} {Phys. Rev. Lett.}\ }\textbf {\bibinfo {volume} {117}},\
  \bibinfo {pages} {267002} (\bibinfo {year} {2016})}\BibitemShut {NoStop}%
\bibitem [{\citenamefont {Dornheim}\ \emph
  {et~al.}(2015{\natexlab{a}})\citenamefont {Dornheim}, \citenamefont {Groth},
  \citenamefont {Filinov},\ and\ \citenamefont
  {Bonitz}}]{dornheim2015permutation}%
  \BibitemOpen
  \bibfield  {author} {\bibinfo {author} {\bibfnamefont {T.}~\bibnamefont
  {Dornheim}}, \bibinfo {author} {\bibfnamefont {S.}~\bibnamefont {Groth}},
  \bibinfo {author} {\bibfnamefont {A.}~\bibnamefont {Filinov}},\ and\ \bibinfo
  {author} {\bibfnamefont {M.}~\bibnamefont {Bonitz}},\ }\bibfield  {title}
  {\bibinfo {title} {Permutation blocking path integral monte carlo: a highly
  efficient approach to the simulation of strongly degenerate non-ideal
  fermions},\ }\href@noop {} {\bibfield  {journal} {\bibinfo  {journal} {New
  Journal of Physics}\ }\textbf {\bibinfo {volume} {17}},\ \bibinfo {pages}
  {073017} (\bibinfo {year} {2015}{\natexlab{a}})}\BibitemShut {NoStop}%
\bibitem [{\citenamefont {Hirshberg}\ \emph {et~al.}(2020)\citenamefont
  {Hirshberg}, \citenamefont {Invernizzi},\ and\ \citenamefont
  {Parrinello}}]{hirshberg2020path}%
  \BibitemOpen
  \bibfield  {author} {\bibinfo {author} {\bibfnamefont {B.}~\bibnamefont
  {Hirshberg}}, \bibinfo {author} {\bibfnamefont {M.}~\bibnamefont
  {Invernizzi}},\ and\ \bibinfo {author} {\bibfnamefont {M.}~\bibnamefont
  {Parrinello}},\ }\bibfield  {title} {\bibinfo {title} {Path integral
  molecular dynamics for fermions: Alleviating the sign problem with the
  bogoliubov inequality},\ }\href {https://doi.org/10.1063/5.0008720}
  {\bibfield  {journal} {\bibinfo  {journal} {The Journal of Chemical Physics}\
  }\textbf {\bibinfo {volume} {152}},\ \bibinfo {pages} {171102} (\bibinfo
  {year} {2020})}\BibitemShut {NoStop}%
\bibitem [{\citenamefont {Dornheim}\ \emph {et~al.}(2020)\citenamefont
  {Dornheim}, \citenamefont {Invernizzi}, \citenamefont {Vorberger},\ and\
  \citenamefont {Hirshberg}}]{dornheim2020attenuating}%
  \BibitemOpen
  \bibfield  {author} {\bibinfo {author} {\bibfnamefont {T.}~\bibnamefont
  {Dornheim}}, \bibinfo {author} {\bibfnamefont {M.}~\bibnamefont
  {Invernizzi}}, \bibinfo {author} {\bibfnamefont {J.}~\bibnamefont
  {Vorberger}},\ and\ \bibinfo {author} {\bibfnamefont {B.}~\bibnamefont
  {Hirshberg}},\ }\bibfield  {title} {\bibinfo {title} {Attenuating the fermion
  sign problem in path integral monte carlo simulations using the bogoliubov
  inequality and thermodynamic integration},\ }\href
  {https://doi.org/10.1063/5.0030760} {\bibfield  {journal} {\bibinfo
  {journal} {The Journal of Chemical Physics}\ }\textbf {\bibinfo {volume}
  {153}},\ \bibinfo {pages} {234104} (\bibinfo {year} {2020})}\BibitemShut
  {NoStop}%
\bibitem [{\citenamefont {Vitenburgs}\ and\ \citenamefont
  {Frost}(2026)}]{vitenburgs2026}%
  \BibitemOpen
  \bibfield  {author} {\bibinfo {author} {\bibfnamefont {I.}~\bibnamefont
  {Vitenburgs}}\ and\ \bibinfo {author} {\bibfnamefont {J.~M.}\ \bibnamefont
  {Frost}},\ }\bibfield  {title} {\bibinfo {title} {Hydrodynamic backflow for
  easing the fermion sign in finite-temperature electron path integral
  simulations},\ }\href@noop {} {\bibfield  {journal} {\bibinfo  {journal}
  {arXiv preprint arXiv:2604.01963}\ } (\bibinfo {year} {2026})}\BibitemShut
  {NoStop}%
\bibitem [{\citenamefont {Bonitz}\ \emph {et~al.}(2024)\citenamefont {Bonitz},
  \citenamefont {Vorberger}, \citenamefont {Bethkenhagen}, \citenamefont
  {Böhme}, \citenamefont {Ceperley} \emph {et~al.}}]{bonitz2024toward}%
  \BibitemOpen
  \bibfield  {author} {\bibinfo {author} {\bibfnamefont {M.}~\bibnamefont
  {Bonitz}}, \bibinfo {author} {\bibfnamefont {J.}~\bibnamefont {Vorberger}},
  \bibinfo {author} {\bibfnamefont {M.}~\bibnamefont {Bethkenhagen}}, \bibinfo
  {author} {\bibfnamefont {M.~P.}\ \bibnamefont {Böhme}}, \bibinfo {author}
  {\bibfnamefont {D.~M.}\ \bibnamefont {Ceperley}}, \emph {et~al.},\ }\bibfield
   {title} {\bibinfo {title} {Toward first principles-based simulations of
  dense hydrogen},\ }\href@noop {} {\bibfield  {journal} {\bibinfo  {journal}
  {Physics of Plasmas}\ }\textbf {\bibinfo {volume} {31}},\ \bibinfo {pages}
  {110501} (\bibinfo {year} {2024})}\BibitemShut {NoStop}%
\bibitem [{\citenamefont {Vorberger}\ \emph {et~al.}(2025)\citenamefont
  {Vorberger}, \citenamefont {Graziani}, \citenamefont {Riley}, \citenamefont
  {Baczewski}, \citenamefont {Baraffe}, \citenamefont {Bethkenhagen},
  \citenamefont {Blouin}, \citenamefont {B{\"o}hme}, \citenamefont {Bonitz},
  \citenamefont {Bussmann} \emph {et~al.}}]{vorberger2025roadmap}%
  \BibitemOpen
  \bibfield  {author} {\bibinfo {author} {\bibfnamefont {J.}~\bibnamefont
  {Vorberger}}, \bibinfo {author} {\bibfnamefont {F.}~\bibnamefont {Graziani}},
  \bibinfo {author} {\bibfnamefont {D.}~\bibnamefont {Riley}}, \bibinfo
  {author} {\bibfnamefont {A.~D.}\ \bibnamefont {Baczewski}}, \bibinfo {author}
  {\bibfnamefont {I.}~\bibnamefont {Baraffe}}, \bibinfo {author} {\bibfnamefont
  {M.}~\bibnamefont {Bethkenhagen}}, \bibinfo {author} {\bibfnamefont
  {S.}~\bibnamefont {Blouin}}, \bibinfo {author} {\bibfnamefont {M.~P.}\
  \bibnamefont {B{\"o}hme}}, \bibinfo {author} {\bibfnamefont {M.}~\bibnamefont
  {Bonitz}}, \bibinfo {author} {\bibfnamefont {M.}~\bibnamefont {Bussmann}},
  \emph {et~al.},\ }\bibfield  {title} {\bibinfo {title} {Roadmap for warm
  dense matter physics},\ }\href@noop {} {\bibfield  {journal} {\bibinfo
  {journal} {arXiv preprint arXiv:2505.02494}\ } (\bibinfo {year}
  {2025})}\BibitemShut {NoStop}%
\bibitem [{\citenamefont {Bonitz}\ \emph {et~al.}(2026)\citenamefont {Bonitz},
  \citenamefont {K{\"a}hlert}, \citenamefont {Krimans}, \citenamefont {Makait},
  \citenamefont {Hamann}, \citenamefont {Vorberger}, \citenamefont
  {Moldabekov}, \citenamefont {Hu}, \citenamefont {Karasiev}, \citenamefont
  {Kraus} \emph {et~al.}}]{bonitz2026}%
  \BibitemOpen
  \bibfield  {author} {\bibinfo {author} {\bibfnamefont {M.}~\bibnamefont
  {Bonitz}}, \bibinfo {author} {\bibfnamefont {H.}~\bibnamefont {K{\"a}hlert}},
  \bibinfo {author} {\bibfnamefont {D.}~\bibnamefont {Krimans}}, \bibinfo
  {author} {\bibfnamefont {C.}~\bibnamefont {Makait}}, \bibinfo {author}
  {\bibfnamefont {P.}~\bibnamefont {Hamann}}, \bibinfo {author} {\bibfnamefont
  {J.}~\bibnamefont {Vorberger}}, \bibinfo {author} {\bibfnamefont
  {Z.}~\bibnamefont {Moldabekov}}, \bibinfo {author} {\bibfnamefont
  {S.}~\bibnamefont {Hu}}, \bibinfo {author} {\bibfnamefont {V.}~\bibnamefont
  {Karasiev}}, \bibinfo {author} {\bibfnamefont {D.}~\bibnamefont {Kraus}},
  \emph {et~al.},\ }\bibfield  {title} {\bibinfo {title} {Quantum effects in
  plasmas},\ }\href@noop {} {\bibfield  {journal} {\bibinfo  {journal} {arXiv
  preprint arXiv:2604.03757}\ } (\bibinfo {year} {2026})}\BibitemShut {NoStop}%
\bibitem [{\citenamefont {Zhang}\ \emph {et~al.}(1997)\citenamefont {Zhang},
  \citenamefont {Carlson},\ and\ \citenamefont
  {Gubernatis}}]{zhang1997constrained}%
  \BibitemOpen
  \bibfield  {author} {\bibinfo {author} {\bibfnamefont {S.}~\bibnamefont
  {Zhang}}, \bibinfo {author} {\bibfnamefont {J.}~\bibnamefont {Carlson}},\
  and\ \bibinfo {author} {\bibfnamefont {J.~E.}\ \bibnamefont {Gubernatis}},\
  }\bibfield  {title} {\bibinfo {title} {Constrained path monte carlo method
  for fermion ground states},\ }\href@noop {} {\bibfield  {journal} {\bibinfo
  {journal} {Physical Review B}\ }\textbf {\bibinfo {volume} {55}},\ \bibinfo
  {pages} {7464} (\bibinfo {year} {1997})}\BibitemShut {NoStop}%
\bibitem [{\citenamefont {Xiong}\ and\ \citenamefont
  {Xiong}(2022)}]{xiong2022thermodynamic}%
  \BibitemOpen
  \bibfield  {author} {\bibinfo {author} {\bibfnamefont {Y.}~\bibnamefont
  {Xiong}}\ and\ \bibinfo {author} {\bibfnamefont {H.}~\bibnamefont {Xiong}},\
  }\bibfield  {title} {\bibinfo {title} {On the thermodynamic properties of
  fictitious identical particles and the application to fermion
  sign problem},\ }\href {https://doi.org/10.1063/5.0106067} {\bibfield
  {journal} {\bibinfo  {journal} {The Journal of Chemical Physics}\ }\textbf
  {\bibinfo {volume} {157}},\ \bibinfo {pages} {094112} (\bibinfo {year}
  {2022})}\BibitemShut {NoStop}%
\bibitem [{\citenamefont {Xiong}\ and\ \citenamefont
  {Xiong}(2023)}]{xiong2023thermodynamics}%
  \BibitemOpen
  \bibfield  {author} {\bibinfo {author} {\bibfnamefont {Y.}~\bibnamefont
  {Xiong}}\ and\ \bibinfo {author} {\bibfnamefont {H.}~\bibnamefont {Xiong}},\
  }\bibfield  {title} {\bibinfo {title} {Thermodynamics of fermions at any
  temperature based on parametrized partition function},\ }\href@noop {}
  {\bibfield  {journal} {\bibinfo  {journal} {Physical Review E}\ }\textbf
  {\bibinfo {volume} {107}},\ \bibinfo {pages} {055308} (\bibinfo {year}
  {2023})}\BibitemShut {NoStop}%
\bibitem [{\citenamefont {Dornheim}\ \emph {et~al.}(2023)\citenamefont
  {Dornheim}, \citenamefont {Tolias}, \citenamefont {Groth}, \citenamefont
  {Moldabekov}, \citenamefont {Vorberger},\ and\ \citenamefont
  {Hirshberg}}]{dornheim2023fermionic}%
  \BibitemOpen
  \bibfield  {author} {\bibinfo {author} {\bibfnamefont {T.}~\bibnamefont
  {Dornheim}}, \bibinfo {author} {\bibfnamefont {P.}~\bibnamefont {Tolias}},
  \bibinfo {author} {\bibfnamefont {S.}~\bibnamefont {Groth}}, \bibinfo
  {author} {\bibfnamefont {Z.~A.}\ \bibnamefont {Moldabekov}}, \bibinfo
  {author} {\bibfnamefont {J.}~\bibnamefont {Vorberger}},\ and\ \bibinfo
  {author} {\bibfnamefont {B.}~\bibnamefont {Hirshberg}},\ }\bibfield  {title}
  {\bibinfo {title} {Fermionic physics from ab initio path integral monte carlo
  simulations of fictitious identical particles},\ }\href
  {https://doi.org/10.1063/5.0171930} {\bibfield  {journal} {\bibinfo
  {journal} {The Journal of Chemical Physics}\ }\textbf {\bibinfo {volume}
  {159}},\ \bibinfo {pages} {164113} (\bibinfo {year} {2023})}\BibitemShut
  {NoStop}%
\bibitem [{\citenamefont {Dornheim}\ \emph
  {et~al.}(2025{\natexlab{a}})\citenamefont {Dornheim}, \citenamefont
  {D{\"o}ppner}, \citenamefont {Tolias}, \citenamefont {B{\"o}hme},
  \citenamefont {Fletcher}, \citenamefont {Gawne}, \citenamefont {Graziani},
  \citenamefont {Kraus}, \citenamefont {MacDonald}, \citenamefont {Moldabekov}
  \emph {et~al.}}]{dornheim2025unraveling}%
  \BibitemOpen
  \bibfield  {author} {\bibinfo {author} {\bibfnamefont {T.}~\bibnamefont
  {Dornheim}}, \bibinfo {author} {\bibfnamefont {T.}~\bibnamefont
  {D{\"o}ppner}}, \bibinfo {author} {\bibfnamefont {P.}~\bibnamefont {Tolias}},
  \bibinfo {author} {\bibfnamefont {M.~P.}\ \bibnamefont {B{\"o}hme}}, \bibinfo
  {author} {\bibfnamefont {L.~B.}\ \bibnamefont {Fletcher}}, \bibinfo {author}
  {\bibfnamefont {T.}~\bibnamefont {Gawne}}, \bibinfo {author} {\bibfnamefont
  {F.~R.}\ \bibnamefont {Graziani}}, \bibinfo {author} {\bibfnamefont
  {D.}~\bibnamefont {Kraus}}, \bibinfo {author} {\bibfnamefont {M.~J.}\
  \bibnamefont {MacDonald}}, \bibinfo {author} {\bibfnamefont {Z.~A.}\
  \bibnamefont {Moldabekov}}, \emph {et~al.},\ }\bibfield  {title} {\bibinfo
  {title} {Unraveling electronic correlations in warm dense quantum plasmas},\
  }\href@noop {} {\bibfield  {journal} {\bibinfo  {journal} {Nature
  Communications}\ }\textbf {\bibinfo {volume} {16}},\ \bibinfo {pages} {5103}
  (\bibinfo {year} {2025}{\natexlab{a}})}\BibitemShut {NoStop}%
\bibitem [{\citenamefont {Dornheim}\ \emph
  {et~al.}(2025{\natexlab{b}})\citenamefont {Dornheim}, \citenamefont {Robles},
  \citenamefont {Hamann}, \citenamefont {Chuna}, \citenamefont {Svensson},
  \citenamefont {Schwalbe}, \citenamefont {Moldabekov}, \citenamefont
  {Tolias},\ and\ \citenamefont {Vorberger}}]{dornheim2025taylor}%
  \BibitemOpen
  \bibfield  {author} {\bibinfo {author} {\bibfnamefont {T.}~\bibnamefont
  {Dornheim}}, \bibinfo {author} {\bibfnamefont {A.~B.}\ \bibnamefont
  {Robles}}, \bibinfo {author} {\bibfnamefont {P.}~\bibnamefont {Hamann}},
  \bibinfo {author} {\bibfnamefont {T.~M.}\ \bibnamefont {Chuna}}, \bibinfo
  {author} {\bibfnamefont {P.}~\bibnamefont {Svensson}}, \bibinfo {author}
  {\bibfnamefont {S.}~\bibnamefont {Schwalbe}}, \bibinfo {author}
  {\bibfnamefont {Z.~A.}\ \bibnamefont {Moldabekov}}, \bibinfo {author}
  {\bibfnamefont {P.}~\bibnamefont {Tolias}},\ and\ \bibinfo {author}
  {\bibfnamefont {J.}~\bibnamefont {Vorberger}},\ }\bibfield  {title} {\bibinfo
  {title} {Taylor series perspective on ab initio path integral monte carlo
  simulations with fermi-dirac statistics},\ }\href@noop {} {\bibfield
  {journal} {\bibinfo  {journal} {arXiv preprint arXiv:2509.11317}\ } (\bibinfo
  {year} {2025}{\natexlab{b}})}\BibitemShut {NoStop}%
\bibitem [{\citenamefont {Morresi}\ and\ \citenamefont
  {Garberoglio}(2025)}]{morresi2025normal}%
  \BibitemOpen
  \bibfield  {author} {\bibinfo {author} {\bibfnamefont {T.}~\bibnamefont
  {Morresi}}\ and\ \bibinfo {author} {\bibfnamefont {G.}~\bibnamefont
  {Garberoglio}},\ }\bibfield  {title} {\bibinfo {title} {Normal liquid
  {$\text{He}^3$} studied by path-integral monte carlo with a parametrized
  partition function},\ }\href@noop {} {\bibfield  {journal} {\bibinfo
  {journal} {Physical Review B}\ }\textbf {\bibinfo {volume} {111}},\ \bibinfo
  {pages} {014521} (\bibinfo {year} {2025})}\BibitemShut {NoStop}%
\bibitem [{\citenamefont {Fan}\ \emph {et~al.}(2026)\citenamefont {Fan},
  \citenamefont {Xiao},\ and\ \citenamefont {Deng}}]{fan2025quantum}%
  \BibitemOpen
  \bibfield  {author} {\bibinfo {author} {\bibfnamefont {Z.}~\bibnamefont
  {Fan}}, \bibinfo {author} {\bibfnamefont {T.}~\bibnamefont {Xiao}},\ and\
  \bibinfo {author} {\bibfnamefont {Y.}~\bibnamefont {Deng}},\ }\bibfield
  {title} {\bibinfo {title} {Quantum path-integral method for the
  fictitious-particle hubbard model},\ }\href
  {https://doi.org/10.1103/vqj6-6vyt} {\bibfield  {journal} {\bibinfo
  {journal} {Phys. Rev. B}\ }\textbf {\bibinfo {volume} {113}},\ \bibinfo
  {pages} {115123} (\bibinfo {year} {2026})}\BibitemShut {NoStop}%
\bibitem [{\citenamefont {Xiong}\ and\ \citenamefont
  {Xiong}(2025)}]{xiong2025pseudo}%
  \BibitemOpen
  \bibfield  {author} {\bibinfo {author} {\bibfnamefont {Y.}~\bibnamefont
  {Xiong}}\ and\ \bibinfo {author} {\bibfnamefont {H.}~\bibnamefont {Xiong}},\
  }\bibfield  {title} {\bibinfo {title} {A pseudo-fermion propagator approach
  to the fermion sign problem},\ }\href {https://doi.org/10.1063/5.0296319}
  {\bibfield  {journal} {\bibinfo  {journal} {The Journal of Chemical Physics}\
  }\textbf {\bibinfo {volume} {163}},\ \bibinfo {pages} {174107} (\bibinfo
  {year} {2025})}\BibitemShut {NoStop}%
\bibitem [{\citenamefont {Xiong}\ \emph {et~al.}(2026)\citenamefont {Xiong},
  \citenamefont {Morresi},\ and\ \citenamefont {Xiong}}]{YunuoUEG}%
  \BibitemOpen
  \bibfield  {author} {\bibinfo {author} {\bibfnamefont {Y.}~\bibnamefont
  {Xiong}}, \bibinfo {author} {\bibfnamefont {T.}~\bibnamefont {Morresi}},\
  and\ \bibinfo {author} {\bibfnamefont {H.}~\bibnamefont {Xiong}},\ }\bibfield
   {title} {\bibinfo {title} {Simulation of strongly quantum-degenerate uniform
  electron gas using the pseudo-fermion method},\ }\href@noop {} {\bibfield
  {journal} {\bibinfo  {journal} {arXiv preprint arXiv:2603.28000}\ } (\bibinfo
  {year} {2026})}\BibitemShut {NoStop}%
\bibitem [{\citenamefont {Blankenbecler}\ \emph {et~al.}(1981)\citenamefont
  {Blankenbecler}, \citenamefont {Scalapino},\ and\ \citenamefont
  {Sugar}}]{blankenbecler1981monte}%
  \BibitemOpen
  \bibfield  {author} {\bibinfo {author} {\bibfnamefont {R.}~\bibnamefont
  {Blankenbecler}}, \bibinfo {author} {\bibfnamefont {D.}~\bibnamefont
  {Scalapino}},\ and\ \bibinfo {author} {\bibfnamefont {R.}~\bibnamefont
  {Sugar}},\ }\bibfield  {title} {\bibinfo {title} {Monte carlo calculations of
  coupled boson-fermion systems. {I}},\ }\href@noop {} {\bibfield  {journal}
  {\bibinfo  {journal} {Physical Review D}\ }\textbf {\bibinfo {volume} {24}},\
  \bibinfo {pages} {2278} (\bibinfo {year} {1981})}\BibitemShut {NoStop}%
\bibitem [{\citenamefont {Hirsch}(1985)}]{hirsch1985two}%
  \BibitemOpen
  \bibfield  {author} {\bibinfo {author} {\bibfnamefont {J.~E.}\ \bibnamefont
  {Hirsch}},\ }\bibfield  {title} {\bibinfo {title} {Two-dimensional hubbard
  model: Numerical simulation study},\ }\href@noop {} {\bibfield  {journal}
  {\bibinfo  {journal} {Physical Review B}\ }\textbf {\bibinfo {volume} {31}},\
  \bibinfo {pages} {4403} (\bibinfo {year} {1985})}\BibitemShut {NoStop}%
\bibitem [{\citenamefont {Lin}\ and\ \citenamefont
  {Hirsch}(1987)}]{lin1987two}%
  \BibitemOpen
  \bibfield  {author} {\bibinfo {author} {\bibfnamefont {H.}~\bibnamefont
  {Lin}}\ and\ \bibinfo {author} {\bibfnamefont {J.}~\bibnamefont {Hirsch}},\
  }\bibfield  {title} {\bibinfo {title} {Two-dimensional hubbard model with
  nearest-and next-nearest-neighbor hopping},\ }\href@noop {} {\bibfield
  {journal} {\bibinfo  {journal} {Physical Review B}\ }\textbf {\bibinfo
  {volume} {35}},\ \bibinfo {pages} {3359} (\bibinfo {year}
  {1987})}\BibitemShut {NoStop}%
\bibitem [{\citenamefont {Hirsch}\ and\ \citenamefont
  {Lin}(1988)}]{hirsch1988pairing}%
  \BibitemOpen
  \bibfield  {author} {\bibinfo {author} {\bibfnamefont {J.}~\bibnamefont
  {Hirsch}}\ and\ \bibinfo {author} {\bibfnamefont {H.}~\bibnamefont {Lin}},\
  }\bibfield  {title} {\bibinfo {title} {Pairing in the two-dimensional
  {H}ubbard model: A {M}onte {C}arlo study},\ }\href@noop {} {\bibfield
  {journal} {\bibinfo  {journal} {Physical Review B}\ }\textbf {\bibinfo
  {volume} {37}},\ \bibinfo {pages} {5070} (\bibinfo {year}
  {1988})}\BibitemShut {NoStop}%
\bibitem [{\citenamefont {White}\ \emph {et~al.}(1989)\citenamefont {White},
  \citenamefont {Scalapino}, \citenamefont {Sugar}, \citenamefont {Loh},
  \citenamefont {Gubernatis},\ and\ \citenamefont
  {Scalettar}}]{white1989numerical}%
  \BibitemOpen
  \bibfield  {author} {\bibinfo {author} {\bibfnamefont {S.~R.}\ \bibnamefont
  {White}}, \bibinfo {author} {\bibfnamefont {D.~J.}\ \bibnamefont
  {Scalapino}}, \bibinfo {author} {\bibfnamefont {R.~L.}\ \bibnamefont
  {Sugar}}, \bibinfo {author} {\bibfnamefont {E.}~\bibnamefont {Loh}}, \bibinfo
  {author} {\bibfnamefont {J.~E.}\ \bibnamefont {Gubernatis}},\ and\ \bibinfo
  {author} {\bibfnamefont {R.~T.}\ \bibnamefont {Scalettar}},\ }\bibfield
  {title} {\bibinfo {title} {Numerical study of the two-dimensional {H}ubbard
  model},\ }\href@noop {} {\bibfield  {journal} {\bibinfo  {journal} {Physical
  Review B}\ }\textbf {\bibinfo {volume} {40}},\ \bibinfo {pages} {506}
  (\bibinfo {year} {1989})}\BibitemShut {NoStop}%
\bibitem [{\citenamefont {Sorella}\ \emph {et~al.}(1989)\citenamefont
  {Sorella}, \citenamefont {Baroni}, \citenamefont {Car},\ and\ \citenamefont
  {Parrinello}}]{sorella1989novel}%
  \BibitemOpen
  \bibfield  {author} {\bibinfo {author} {\bibfnamefont {S.}~\bibnamefont
  {Sorella}}, \bibinfo {author} {\bibfnamefont {S.}~\bibnamefont {Baroni}},
  \bibinfo {author} {\bibfnamefont {R.}~\bibnamefont {Car}},\ and\ \bibinfo
  {author} {\bibfnamefont {M.}~\bibnamefont {Parrinello}},\ }\bibfield  {title}
  {\bibinfo {title} {A novel technique for the simulation of interacting
  fermion systems},\ }\href@noop {} {\bibfield  {journal} {\bibinfo  {journal}
  {Europhysics Letters}\ }\textbf {\bibinfo {volume} {8}},\ \bibinfo {pages}
  {663} (\bibinfo {year} {1989})}\BibitemShut {NoStop}%
\bibitem [{\citenamefont {Santos}(2003)}]{santos2003introduction}%
  \BibitemOpen
  \bibfield  {author} {\bibinfo {author} {\bibfnamefont {R.~R.~d.}\
  \bibnamefont {Santos}},\ }\bibfield  {title} {\bibinfo {title} {Introduction
  to quantum monte carlo simulations for fermionic systems},\ }\href@noop {}
  {\bibfield  {journal} {\bibinfo  {journal} {Brazilian Journal of Physics}\
  }\textbf {\bibinfo {volume} {33}},\ \bibinfo {pages} {36} (\bibinfo {year}
  {2003})}\BibitemShut {NoStop}%
\bibitem [{\citenamefont {Huang}\ \emph {et~al.}(2017)\citenamefont {Huang},
  \citenamefont {Mendl}, \citenamefont {Liu}, \citenamefont {Johnston},
  \citenamefont {Jiang}, \citenamefont {Moritz},\ and\ \citenamefont
  {Devereaux}}]{huang2017numerical}%
  \BibitemOpen
  \bibfield  {author} {\bibinfo {author} {\bibfnamefont {E.~W.}\ \bibnamefont
  {Huang}}, \bibinfo {author} {\bibfnamefont {C.~B.}\ \bibnamefont {Mendl}},
  \bibinfo {author} {\bibfnamefont {S.}~\bibnamefont {Liu}}, \bibinfo {author}
  {\bibfnamefont {S.}~\bibnamefont {Johnston}}, \bibinfo {author}
  {\bibfnamefont {H.-C.}\ \bibnamefont {Jiang}}, \bibinfo {author}
  {\bibfnamefont {B.}~\bibnamefont {Moritz}},\ and\ \bibinfo {author}
  {\bibfnamefont {T.~P.}\ \bibnamefont {Devereaux}},\ }\bibfield  {title}
  {\bibinfo {title} {Numerical evidence of fluctuating stripes in the normal
  state of high-{$T_c$} cuprate superconductors},\ }\href@noop {} {\bibfield
  {journal} {\bibinfo  {journal} {Science}\ }\textbf {\bibinfo {volume}
  {358}},\ \bibinfo {pages} {1161} (\bibinfo {year} {2017})}\BibitemShut
  {NoStop}%
\bibitem [{\citenamefont {Assaad}\ \emph {et~al.}(2025)\citenamefont {Assaad},
  \citenamefont {Bercx}, \citenamefont {Goth}, \citenamefont {G{\"o}tz},
  \citenamefont {Hofmann}, \citenamefont {Huffman}, \citenamefont {Liu},
  \citenamefont {Parisen~Toldin}, \citenamefont {Portela},\ and\ \citenamefont
  {Schwab}}]{assaad2025alf}%
  \BibitemOpen
  \bibfield  {author} {\bibinfo {author} {\bibfnamefont {F.~F.}\ \bibnamefont
  {Assaad}}, \bibinfo {author} {\bibfnamefont {M.}~\bibnamefont {Bercx}},
  \bibinfo {author} {\bibfnamefont {F.}~\bibnamefont {Goth}}, \bibinfo {author}
  {\bibfnamefont {A.}~\bibnamefont {G{\"o}tz}}, \bibinfo {author}
  {\bibfnamefont {J.}~\bibnamefont {Hofmann}}, \bibinfo {author} {\bibfnamefont
  {E.}~\bibnamefont {Huffman}}, \bibinfo {author} {\bibfnamefont
  {Z.}~\bibnamefont {Liu}}, \bibinfo {author} {\bibfnamefont {F.}~\bibnamefont
  {Parisen~Toldin}}, \bibinfo {author} {\bibfnamefont {J.}~\bibnamefont
  {Portela}},\ and\ \bibinfo {author} {\bibfnamefont {J.}~\bibnamefont
  {Schwab}},\ }\bibfield  {title} {\bibinfo {title} {The alf (algorithms for
  lattice fermions) project release 2.4. documentation for the auxiliary-field
  quantum monte carlo code},\ }\href@noop {} {\bibfield  {journal} {\bibinfo
  {journal} {SciPost Physics Codebases}\ ,\ \bibinfo {pages} {1}} (\bibinfo
  {year} {2025})}\BibitemShut {NoStop}%
\bibitem [{\citenamefont {Song}\ \emph {et~al.}(2025)\citenamefont {Song},
  \citenamefont {Deng},\ and\ \citenamefont {He}}]{song2025extended}%
  \BibitemOpen
  \bibfield  {author} {\bibinfo {author} {\bibfnamefont {Y.-F.}\ \bibnamefont
  {Song}}, \bibinfo {author} {\bibfnamefont {Y.}~\bibnamefont {Deng}},\ and\
  \bibinfo {author} {\bibfnamefont {Y.-Y.}\ \bibnamefont {He}},\ }\bibfield
  {title} {\bibinfo {title} {Extended metal-insulator crossover with strong
  antiferromagnetic spin correlation in half-filled 3d hubbard model},\
  }\href@noop {} {\bibfield  {journal} {\bibinfo  {journal} {Physical Review
  Letters}\ }\textbf {\bibinfo {volume} {134}},\ \bibinfo {pages} {016503}
  (\bibinfo {year} {2025})}\BibitemShut {NoStop}%
\bibitem [{\citenamefont {Zhang}\ \emph {et~al.}(1995)\citenamefont {Zhang},
  \citenamefont {Carlson},\ and\ \citenamefont
  {Gubernatis}}]{zhang1995constrained}%
  \BibitemOpen
  \bibfield  {author} {\bibinfo {author} {\bibfnamefont {S.}~\bibnamefont
  {Zhang}}, \bibinfo {author} {\bibfnamefont {J.}~\bibnamefont {Carlson}},\
  and\ \bibinfo {author} {\bibfnamefont {J.~E.}\ \bibnamefont {Gubernatis}},\
  }\bibfield  {title} {\bibinfo {title} {Constrained path quantum monte carlo
  method for fermion ground states},\ }\href@noop {} {\bibfield  {journal}
  {\bibinfo  {journal} {Physical Review Letters}\ }\textbf {\bibinfo {volume}
  {74}},\ \bibinfo {pages} {3652} (\bibinfo {year} {1995})}\BibitemShut
  {NoStop}%
\bibitem [{\citenamefont {Anderson}(1975)}]{anderson1975random}%
  \BibitemOpen
  \bibfield  {author} {\bibinfo {author} {\bibfnamefont {J.~B.}\ \bibnamefont
  {Anderson}},\ }\bibfield  {title} {\bibinfo {title} {A random-walk simulation
  of the schr{\"o}dinger equation: {$H_3^+$}},\ }\href@noop {} {\bibfield
  {journal} {\bibinfo  {journal} {The Journal of Chemical Physics}\ }\textbf
  {\bibinfo {volume} {63}},\ \bibinfo {pages} {1499} (\bibinfo {year}
  {1975})}\BibitemShut {NoStop}%
\bibitem [{\citenamefont {Ceperley}\ and\ \citenamefont
  {Alder}(1980)}]{Ceperley1980}%
  \BibitemOpen
  \bibfield  {author} {\bibinfo {author} {\bibfnamefont {D.~M.}\ \bibnamefont
  {Ceperley}}\ and\ \bibinfo {author} {\bibfnamefont {B.~J.}\ \bibnamefont
  {Alder}},\ }\bibfield  {title} {\bibinfo {title} {Ground state of the
  electron gas by a stochastic method},\ }\href
  {https://doi.org/10.1103/PhysRevLett.45.566} {\bibfield  {journal} {\bibinfo
  {journal} {Phys. Rev. Lett.}\ }\textbf {\bibinfo {volume} {45}},\ \bibinfo
  {pages} {566} (\bibinfo {year} {1980})}\BibitemShut {NoStop}%
\bibitem [{\citenamefont {LeBlanc}\ \emph {et~al.}(2015)\citenamefont
  {LeBlanc}, \citenamefont {Antipov}, \citenamefont {Becca}, \citenamefont
  {Bulik}, \citenamefont {Chan}, \citenamefont {Chung}, \citenamefont {Deng},
  \citenamefont {Ferrero}, \citenamefont {Henderson}, \citenamefont
  {Jim{\'e}nez-Hoyos} \emph {et~al.}}]{leblanc2015solutions}%
  \BibitemOpen
  \bibfield  {author} {\bibinfo {author} {\bibfnamefont {J.~P.}\ \bibnamefont
  {LeBlanc}}, \bibinfo {author} {\bibfnamefont {A.~E.}\ \bibnamefont
  {Antipov}}, \bibinfo {author} {\bibfnamefont {F.}~\bibnamefont {Becca}},
  \bibinfo {author} {\bibfnamefont {I.~W.}\ \bibnamefont {Bulik}}, \bibinfo
  {author} {\bibfnamefont {G.~K.-L.}\ \bibnamefont {Chan}}, \bibinfo {author}
  {\bibfnamefont {C.-M.}\ \bibnamefont {Chung}}, \bibinfo {author}
  {\bibfnamefont {Y.}~\bibnamefont {Deng}}, \bibinfo {author} {\bibfnamefont
  {M.}~\bibnamefont {Ferrero}}, \bibinfo {author} {\bibfnamefont {T.~M.}\
  \bibnamefont {Henderson}}, \bibinfo {author} {\bibfnamefont {C.~A.}\
  \bibnamefont {Jim{\'e}nez-Hoyos}}, \emph {et~al.},\ }\bibfield  {title}
  {\bibinfo {title} {Solutions of the two-dimensional hubbard model: Benchmarks
  and results from a wide range of numerical algorithms},\ }\href@noop {}
  {\bibfield  {journal} {\bibinfo  {journal} {Physical Review X}\ }\textbf
  {\bibinfo {volume} {5}},\ \bibinfo {pages} {041041} (\bibinfo {year}
  {2015})}\BibitemShut {NoStop}%
\bibitem [{\citenamefont {Zheng}\ \emph {et~al.}(2017)\citenamefont {Zheng},
  \citenamefont {Chung}, \citenamefont {Corboz}, \citenamefont {Ehlers},
  \citenamefont {Qin}, \citenamefont {Noack}, \citenamefont {Shi},
  \citenamefont {White}, \citenamefont {Zhang},\ and\ \citenamefont
  {Chan}}]{zheng2017stripe}%
  \BibitemOpen
  \bibfield  {author} {\bibinfo {author} {\bibfnamefont {B.-X.}\ \bibnamefont
  {Zheng}}, \bibinfo {author} {\bibfnamefont {C.-M.}\ \bibnamefont {Chung}},
  \bibinfo {author} {\bibfnamefont {P.}~\bibnamefont {Corboz}}, \bibinfo
  {author} {\bibfnamefont {G.}~\bibnamefont {Ehlers}}, \bibinfo {author}
  {\bibfnamefont {M.-P.}\ \bibnamefont {Qin}}, \bibinfo {author} {\bibfnamefont
  {R.~M.}\ \bibnamefont {Noack}}, \bibinfo {author} {\bibfnamefont
  {H.}~\bibnamefont {Shi}}, \bibinfo {author} {\bibfnamefont {S.~R.}\
  \bibnamefont {White}}, \bibinfo {author} {\bibfnamefont {S.}~\bibnamefont
  {Zhang}},\ and\ \bibinfo {author} {\bibfnamefont {G.~K.-L.}\ \bibnamefont
  {Chan}},\ }\bibfield  {title} {\bibinfo {title} {Stripe order in the
  underdoped region of the two-dimensional hubbard model},\ }\href@noop {}
  {\bibfield  {journal} {\bibinfo  {journal} {Science}\ }\textbf {\bibinfo
  {volume} {358}},\ \bibinfo {pages} {1155} (\bibinfo {year}
  {2017})}\BibitemShut {NoStop}%
\bibitem [{\citenamefont {Imada}\ \emph {et~al.}(1998)\citenamefont {Imada},
  \citenamefont {Fujimori},\ and\ \citenamefont {Tokura}}]{imada1998metal}%
  \BibitemOpen
  \bibfield  {author} {\bibinfo {author} {\bibfnamefont {M.}~\bibnamefont
  {Imada}}, \bibinfo {author} {\bibfnamefont {A.}~\bibnamefont {Fujimori}},\
  and\ \bibinfo {author} {\bibfnamefont {Y.}~\bibnamefont {Tokura}},\
  }\bibfield  {title} {\bibinfo {title} {Metal-insulator transitions},\
  }\href@noop {} {\bibfield  {journal} {\bibinfo  {journal} {Reviews of modern
  physics}\ }\textbf {\bibinfo {volume} {70}},\ \bibinfo {pages} {1039}
  (\bibinfo {year} {1998})}\BibitemShut {NoStop}%
\bibitem [{\citenamefont {Arovas}\ \emph {et~al.}(2022)\citenamefont {Arovas},
  \citenamefont {Berg}, \citenamefont {Kivelson},\ and\ \citenamefont
  {Raghu}}]{arovas2022hubbard}%
  \BibitemOpen
  \bibfield  {author} {\bibinfo {author} {\bibfnamefont {D.~P.}\ \bibnamefont
  {Arovas}}, \bibinfo {author} {\bibfnamefont {E.}~\bibnamefont {Berg}},
  \bibinfo {author} {\bibfnamefont {S.~A.}\ \bibnamefont {Kivelson}},\ and\
  \bibinfo {author} {\bibfnamefont {S.}~\bibnamefont {Raghu}},\ }\bibfield
  {title} {\bibinfo {title} {The hubbard model},\ }\href@noop {} {\bibfield
  {journal} {\bibinfo  {journal} {Annual review of condensed matter physics}\
  }\textbf {\bibinfo {volume} {13}},\ \bibinfo {pages} {239} (\bibinfo {year}
  {2022})}\BibitemShut {NoStop}%
\bibitem [{\citenamefont {Qin}\ \emph {et~al.}(2022)\citenamefont {Qin},
  \citenamefont {Sch{\"a}fer}, \citenamefont {Andergassen}, \citenamefont
  {Corboz},\ and\ \citenamefont {Gull}}]{qin2022hubbard}%
  \BibitemOpen
  \bibfield  {author} {\bibinfo {author} {\bibfnamefont {M.}~\bibnamefont
  {Qin}}, \bibinfo {author} {\bibfnamefont {T.}~\bibnamefont {Sch{\"a}fer}},
  \bibinfo {author} {\bibfnamefont {S.}~\bibnamefont {Andergassen}}, \bibinfo
  {author} {\bibfnamefont {P.}~\bibnamefont {Corboz}},\ and\ \bibinfo {author}
  {\bibfnamefont {E.}~\bibnamefont {Gull}},\ }\bibfield  {title} {\bibinfo
  {title} {The {H}ubbard model: A computational perspective},\ }\href@noop {}
  {\bibfield  {journal} {\bibinfo  {journal} {Annual Review of Condensed Matter
  Physics}\ }\textbf {\bibinfo {volume} {13}},\ \bibinfo {pages} {275}
  (\bibinfo {year} {2022})}\BibitemShut {NoStop}%
\bibitem [{\citenamefont {Verstraete}\ \emph {et~al.}(2008)\citenamefont
  {Verstraete}, \citenamefont {Murg},\ and\ \citenamefont
  {Cirac}}]{verstraete2008matrix}%
  \BibitemOpen
  \bibfield  {author} {\bibinfo {author} {\bibfnamefont {F.}~\bibnamefont
  {Verstraete}}, \bibinfo {author} {\bibfnamefont {V.}~\bibnamefont {Murg}},\
  and\ \bibinfo {author} {\bibfnamefont {J.~I.}\ \bibnamefont {Cirac}},\
  }\bibfield  {title} {\bibinfo {title} {Matrix product states, projected
  entangled pair states, and variational renormalization group methods for
  quantum spin systems},\ }\href@noop {} {\bibfield  {journal} {\bibinfo
  {journal} {Advances in physics}\ }\textbf {\bibinfo {volume} {57}},\ \bibinfo
  {pages} {143} (\bibinfo {year} {2008})}\BibitemShut {NoStop}%
\bibitem [{\citenamefont {Liu}\ \emph {et~al.}(2021)\citenamefont {Liu},
  \citenamefont {Huang}, \citenamefont {Gong},\ and\ \citenamefont
  {Gu}}]{Liu_PRB}%
  \BibitemOpen
  \bibfield  {author} {\bibinfo {author} {\bibfnamefont {W.-Y.}\ \bibnamefont
  {Liu}}, \bibinfo {author} {\bibfnamefont {Y.-Z.}\ \bibnamefont {Huang}},
  \bibinfo {author} {\bibfnamefont {S.-S.}\ \bibnamefont {Gong}},\ and\
  \bibinfo {author} {\bibfnamefont {Z.-C.}\ \bibnamefont {Gu}},\ }\bibfield
  {title} {\bibinfo {title} {Accurate simulation for finite projected entangled
  pair states in two dimensions},\ }\href
  {https://doi.org/10.1103/PhysRevB.103.235155} {\bibfield  {journal} {\bibinfo
   {journal} {Phys. Rev. B}\ }\textbf {\bibinfo {volume} {103}},\ \bibinfo
  {pages} {235155} (\bibinfo {year} {2021})}\BibitemShut {NoStop}%
\bibitem [{\citenamefont {Liu}\ \emph {et~al.}(2025{\natexlab{a}})\citenamefont
  {Liu}, \citenamefont {Zhai}, \citenamefont {Peng}, \citenamefont {Gu},\ and\
  \citenamefont {Chan}}]{WY_Liu}%
  \BibitemOpen
  \bibfield  {author} {\bibinfo {author} {\bibfnamefont {W.-Y.}\ \bibnamefont
  {Liu}}, \bibinfo {author} {\bibfnamefont {H.}~\bibnamefont {Zhai}}, \bibinfo
  {author} {\bibfnamefont {R.}~\bibnamefont {Peng}}, \bibinfo {author}
  {\bibfnamefont {Z.-C.}\ \bibnamefont {Gu}},\ and\ \bibinfo {author}
  {\bibfnamefont {G.~K.-L.}\ \bibnamefont {Chan}},\ }\bibfield  {title}
  {\bibinfo {title} {Accurate simulation of the hubbard model with finite
  fermionic projected entangled pair states},\ }\href
  {https://doi.org/10.1103/r4q9-4yvj} {\bibfield  {journal} {\bibinfo
  {journal} {Phys. Rev. Lett.}\ }\textbf {\bibinfo {volume} {134}},\ \bibinfo
  {pages} {256502} (\bibinfo {year} {2025}{\natexlab{a}})}\BibitemShut
  {NoStop}%
\bibitem [{\citenamefont {Liu}\ \emph {et~al.}(2025{\natexlab{b}})\citenamefont
  {Liu}, \citenamefont {Zhai}, \citenamefont {Peng}, \citenamefont {Gu},\ and\
  \citenamefont {Chan}}]{liu2025accurate}%
  \BibitemOpen
  \bibfield  {author} {\bibinfo {author} {\bibfnamefont {W.-Y.}\ \bibnamefont
  {Liu}}, \bibinfo {author} {\bibfnamefont {H.}~\bibnamefont {Zhai}}, \bibinfo
  {author} {\bibfnamefont {R.}~\bibnamefont {Peng}}, \bibinfo {author}
  {\bibfnamefont {Z.-C.}\ \bibnamefont {Gu}},\ and\ \bibinfo {author}
  {\bibfnamefont {G.~K.-L.}\ \bibnamefont {Chan}},\ }\bibfield  {title}
  {\bibinfo {title} {Accurate simulation of the hubbard model with finite
  fermionic projected entangled pair states},\ }\href@noop {} {\bibfield
  {journal} {\bibinfo  {journal} {Physical Review Letters}\ }\textbf {\bibinfo
  {volume} {134}},\ \bibinfo {pages} {256502} (\bibinfo {year}
  {2025}{\natexlab{b}})}\BibitemShut {NoStop}%
\bibitem [{\citenamefont {Knizia}\ and\ \citenamefont
  {Chan}(2012)}]{knizia2012density}%
  \BibitemOpen
  \bibfield  {author} {\bibinfo {author} {\bibfnamefont {G.}~\bibnamefont
  {Knizia}}\ and\ \bibinfo {author} {\bibfnamefont {G.~K.-L.}\ \bibnamefont
  {Chan}},\ }\bibfield  {title} {\bibinfo {title} {Density matrix embedding: A
  simple alternative to dynamical mean-field theory},\ }\href@noop {}
  {\bibfield  {journal} {\bibinfo  {journal} {Physical Review Letters}\
  }\textbf {\bibinfo {volume} {109}},\ \bibinfo {pages} {186404} (\bibinfo
  {year} {2012})}\BibitemShut {NoStop}%
\bibitem [{\citenamefont {Fahy}\ and\ \citenamefont
  {Hamann}(1990)}]{fahy1990positive}%
  \BibitemOpen
  \bibfield  {author} {\bibinfo {author} {\bibfnamefont {S.}~\bibnamefont
  {Fahy}}\ and\ \bibinfo {author} {\bibfnamefont {D.}~\bibnamefont {Hamann}},\
  }\bibfield  {title} {\bibinfo {title} {Positive-projection monte carlo
  simulation: A new variational approach to strongly interacting fermion
  systems},\ }\href@noop {} {\bibfield  {journal} {\bibinfo  {journal}
  {Physical Review Letters}\ }\textbf {\bibinfo {volume} {65}},\ \bibinfo
  {pages} {3437} (\bibinfo {year} {1990})}\BibitemShut {NoStop}%
\bibitem [{\citenamefont {Zhang}(1999)}]{zhang1999finite}%
  \BibitemOpen
  \bibfield  {author} {\bibinfo {author} {\bibfnamefont {S.}~\bibnamefont
  {Zhang}},\ }\bibfield  {title} {\bibinfo {title} {Finite-temperature monte
  carlo calculations for systems with fermions},\ }\href@noop {} {\bibfield
  {journal} {\bibinfo  {journal} {Physical Review Letters}\ }\textbf {\bibinfo
  {volume} {83}},\ \bibinfo {pages} {2777} (\bibinfo {year}
  {1999})}\BibitemShut {NoStop}%
\bibitem [{\citenamefont {Zhang}\ and\ \citenamefont
  {Krakauer}(2003)}]{zhang2003quantum}%
  \BibitemOpen
  \bibfield  {author} {\bibinfo {author} {\bibfnamefont {S.}~\bibnamefont
  {Zhang}}\ and\ \bibinfo {author} {\bibfnamefont {H.}~\bibnamefont
  {Krakauer}},\ }\bibfield  {title} {\bibinfo {title} {Quantum monte carlo
  method using phase-free random walks with slater determinants},\ }\href@noop
  {} {\bibfield  {journal} {\bibinfo  {journal} {Physical Review Letters}\
  }\textbf {\bibinfo {volume} {90}},\ \bibinfo {pages} {136401} (\bibinfo
  {year} {2003})}\BibitemShut {NoStop}%
\bibitem [{\citenamefont {Dagotto}\ \emph {et~al.}(1992)\citenamefont
  {Dagotto}, \citenamefont {Moreo}, \citenamefont {Ortolani}, \citenamefont
  {Poilblanc},\ and\ \citenamefont {Riera}}]{dagotto1992static}%
  \BibitemOpen
  \bibfield  {author} {\bibinfo {author} {\bibfnamefont {E.}~\bibnamefont
  {Dagotto}}, \bibinfo {author} {\bibfnamefont {A.}~\bibnamefont {Moreo}},
  \bibinfo {author} {\bibfnamefont {F.}~\bibnamefont {Ortolani}}, \bibinfo
  {author} {\bibfnamefont {D.}~\bibnamefont {Poilblanc}},\ and\ \bibinfo
  {author} {\bibfnamefont {J.}~\bibnamefont {Riera}},\ }\bibfield  {title}
  {\bibinfo {title} {Static and dynamical properties of doped hubbard
  clusters},\ }\href@noop {} {\bibfield  {journal} {\bibinfo  {journal}
  {Physical Review B}\ }\textbf {\bibinfo {volume} {45}},\ \bibinfo {pages}
  {10741} (\bibinfo {year} {1992})}\BibitemShut {NoStop}%
\bibitem [{\citenamefont {Liang}(2025)}]{liang2025investigating}%
  \BibitemOpen
  \bibfield  {author} {\bibinfo {author} {\bibfnamefont {X.}~\bibnamefont
  {Liang}},\ }\bibfield  {title} {\bibinfo {title} {Investigating the
  fermi-hubbard model by the tensor-backflow method},\ }\href@noop {}
  {\bibfield  {journal} {\bibinfo  {journal} {arXiv preprint arXiv:2507.01856}\
  } (\bibinfo {year} {2025})}\BibitemShut {NoStop}%
\bibitem [{\citenamefont {Zhou}\ \emph {et~al.}(2024)\citenamefont {Zhou},
  \citenamefont {Zhou},\ and\ \citenamefont {Liang}}]{zhou2024solving}%
  \BibitemOpen
  \bibfield  {author} {\bibinfo {author} {\bibfnamefont {Y.-T.}\ \bibnamefont
  {Zhou}}, \bibinfo {author} {\bibfnamefont {Z.-W.}\ \bibnamefont {Zhou}},\
  and\ \bibinfo {author} {\bibfnamefont {X.}~\bibnamefont {Liang}},\ }\bibfield
   {title} {\bibinfo {title} {Solving fermi-hubbard-type models by tensor
  representations of backflow corrections},\ }\href@noop {} {\bibfield
  {journal} {\bibinfo  {journal} {Physical Review B}\ }\textbf {\bibinfo
  {volume} {109}},\ \bibinfo {pages} {245107} (\bibinfo {year}
  {2024})}\BibitemShut {NoStop}%
\bibitem [{\citenamefont {Marino}\ \emph {et~al.}(2022)\citenamefont {Marino},
  \citenamefont {Becca},\ and\ \citenamefont {Tocchio}}]{marino2022stripes}%
  \BibitemOpen
  \bibfield  {author} {\bibinfo {author} {\bibfnamefont {V.}~\bibnamefont
  {Marino}}, \bibinfo {author} {\bibfnamefont {F.}~\bibnamefont {Becca}},\ and\
  \bibinfo {author} {\bibfnamefont {L.~F.}\ \bibnamefont {Tocchio}},\
  }\bibfield  {title} {\bibinfo {title} {Stripes in the extended t-t' hubbard
  model: A variational monte carlo analysis},\ }\href@noop {} {\bibfield
  {journal} {\bibinfo  {journal} {SciPost Physics}\ }\textbf {\bibinfo {volume}
  {12}},\ \bibinfo {pages} {180} (\bibinfo {year} {2022})}\BibitemShut
  {NoStop}%
\bibitem [{\citenamefont {De~Raedt}\ and\ \citenamefont
  {Lagendijk}(1981)}]{Raedt1981}%
  \BibitemOpen
  \bibfield  {author} {\bibinfo {author} {\bibfnamefont {H.}~\bibnamefont
  {De~Raedt}}\ and\ \bibinfo {author} {\bibfnamefont {A.}~\bibnamefont
  {Lagendijk}},\ }\bibfield  {title} {\bibinfo {title} {Monte carlo calculation
  of the thermodynamic properties of a quantum model: A one-dimensional fermion
  lattice model},\ }\href {https://doi.org/10.1103/PhysRevLett.46.77}
  {\bibfield  {journal} {\bibinfo  {journal} {Phys. Rev. Lett.}\ }\textbf
  {\bibinfo {volume} {46}},\ \bibinfo {pages} {77} (\bibinfo {year}
  {1981})}\BibitemShut {NoStop}%
\bibitem [{\citenamefont {Dornheim}\ \emph
  {et~al.}(2015{\natexlab{b}})\citenamefont {Dornheim}, \citenamefont {Groth},
  \citenamefont {Filinov},\ and\ \citenamefont {Bonitz}}]{Dornheim_2015}%
  \BibitemOpen
  \bibfield  {author} {\bibinfo {author} {\bibfnamefont {T.}~\bibnamefont
  {Dornheim}}, \bibinfo {author} {\bibfnamefont {S.}~\bibnamefont {Groth}},
  \bibinfo {author} {\bibfnamefont {A.}~\bibnamefont {Filinov}},\ and\ \bibinfo
  {author} {\bibfnamefont {M.}~\bibnamefont {Bonitz}},\ }\bibfield  {title}
  {\bibinfo {title} {Permutation blocking path integral monte carlo: a highly
  efficient approach to the simulation of strongly degenerate non-ideal
  fermions},\ }\href {https://doi.org/10.1088/1367-2630/17/7/073017} {\bibfield
   {journal} {\bibinfo  {journal} {New Journal of Physics}\ }\textbf {\bibinfo
  {volume} {17}},\ \bibinfo {pages} {073017} (\bibinfo {year}
  {2015}{\natexlab{b}})}\BibitemShut {NoStop}%
\bibitem [{\citenamefont {Chin}(2024)}]{Chin2024}%
  \BibitemOpen
  \bibfield  {author} {\bibinfo {author} {\bibfnamefont {S.~A.}\ \bibnamefont
  {Chin}},\ }\bibfield  {title} {\bibinfo {title} {Simple proof that there is
  no sign problem in path integral monte carlo simulations of fermions in one
  dimension},\ }\href {https://doi.org/10.1103/PhysRevE.109.065312} {\bibfield
  {journal} {\bibinfo  {journal} {Phys. Rev. E}\ }\textbf {\bibinfo {volume}
  {109}},\ \bibinfo {pages} {065312} (\bibinfo {year} {2024})}\BibitemShut
  {NoStop}%
\bibitem [{\citenamefont {de~Jongh}\ \emph {et~al.}(2025)\citenamefont
  {de~Jongh}, \citenamefont {Verstraten}, \citenamefont {Dixmerias},
  \citenamefont {Daix}, \citenamefont {Peaudecerf},\ and\ \citenamefont
  {Yefsah}}]{deJongh2025}%
  \BibitemOpen
  \bibfield  {author} {\bibinfo {author} {\bibfnamefont {T.}~\bibnamefont
  {de~Jongh}}, \bibinfo {author} {\bibfnamefont {J.}~\bibnamefont
  {Verstraten}}, \bibinfo {author} {\bibfnamefont {M.}~\bibnamefont
  {Dixmerias}}, \bibinfo {author} {\bibfnamefont {C.}~\bibnamefont {Daix}},
  \bibinfo {author} {\bibfnamefont {B.}~\bibnamefont {Peaudecerf}},\ and\
  \bibinfo {author} {\bibfnamefont {T.}~\bibnamefont {Yefsah}},\ }\bibfield
  {title} {\bibinfo {title} {Quantum gas microscopy of fermions in the
  continuum},\ }\href {https://doi.org/10.1103/PhysRevLett.134.183403}
  {\bibfield  {journal} {\bibinfo  {journal} {Physical Review Letters}\
  }\textbf {\bibinfo {volume} {134}},\ \bibinfo {pages} {183403} (\bibinfo
  {year} {2025})}\BibitemShut {NoStop}%
\bibitem [{\citenamefont {Yao}\ \emph {et~al.}(2025)\citenamefont {Yao},
  \citenamefont {Chi}, \citenamefont {Wang}, \citenamefont {Fletcher},\ and\
  \citenamefont {Zwierlein}}]{Yao2025}%
  \BibitemOpen
  \bibfield  {author} {\bibinfo {author} {\bibfnamefont {R.}~\bibnamefont
  {Yao}}, \bibinfo {author} {\bibfnamefont {S.}~\bibnamefont {Chi}}, \bibinfo
  {author} {\bibfnamefont {M.}~\bibnamefont {Wang}}, \bibinfo {author}
  {\bibfnamefont {R.~J.}\ \bibnamefont {Fletcher}},\ and\ \bibinfo {author}
  {\bibfnamefont {M.}~\bibnamefont {Zwierlein}},\ }\bibfield  {title} {\bibinfo
  {title} {Measuring pair correlations in bose and fermi gases via
  atom-resolved microscopy},\ }\href
  {https://doi.org/10.1103/PhysRevLett.134.183402} {\bibfield  {journal}
  {\bibinfo  {journal} {Physical Review Letters}\ }\textbf {\bibinfo {volume}
  {134}},\ \bibinfo {pages} {183402} (\bibinfo {year} {2025})}\BibitemShut
  {NoStop}%
\bibitem [{\citenamefont {F{\"o}lling}\ \emph {et~al.}(2005)\citenamefont
  {F{\"o}lling}, \citenamefont {Gerbier}, \citenamefont {Widera}, \citenamefont
  {Mandel}, \citenamefont {Gericke},\ and\ \citenamefont
  {Bloch}}]{folling2005spatial}%
  \BibitemOpen
  \bibfield  {author} {\bibinfo {author} {\bibfnamefont {S.}~\bibnamefont
  {F{\"o}lling}}, \bibinfo {author} {\bibfnamefont {F.}~\bibnamefont
  {Gerbier}}, \bibinfo {author} {\bibfnamefont {A.}~\bibnamefont {Widera}},
  \bibinfo {author} {\bibfnamefont {O.}~\bibnamefont {Mandel}}, \bibinfo
  {author} {\bibfnamefont {T.}~\bibnamefont {Gericke}},\ and\ \bibinfo {author}
  {\bibfnamefont {I.}~\bibnamefont {Bloch}},\ }\bibfield  {title} {\bibinfo
  {title} {Spatial quantum noise interferometry in expanding ultracold atom
  clouds},\ }\href@noop {} {\bibfield  {journal} {\bibinfo  {journal} {Nature}\
  }\textbf {\bibinfo {volume} {434}},\ \bibinfo {pages} {481} (\bibinfo {year}
  {2005})}\BibitemShut {NoStop}%
\bibitem [{\citenamefont {Shen}\ \emph {et~al.}(2020)\citenamefont {Shen},
  \citenamefont {Liu}, \citenamefont {Yu},\ and\ \citenamefont
  {Rubenstein}}]{shen2020finite}%
  \BibitemOpen
  \bibfield  {author} {\bibinfo {author} {\bibfnamefont {T.}~\bibnamefont
  {Shen}}, \bibinfo {author} {\bibfnamefont {Y.}~\bibnamefont {Liu}}, \bibinfo
  {author} {\bibfnamefont {Y.}~\bibnamefont {Yu}},\ and\ \bibinfo {author}
  {\bibfnamefont {B.~M.}\ \bibnamefont {Rubenstein}},\ }\bibfield  {title}
  {\bibinfo {title} {Finite temperature auxiliary field quantum monte carlo in
  the canonical ensemble},\ }\href {https://doi.org/10.1063/5.0026606}
  {\bibfield  {journal} {\bibinfo  {journal} {The Journal of Chemical Physics}\
  }\textbf {\bibinfo {volume} {153}},\ \bibinfo {pages} {204108} (\bibinfo
  {year} {2020})}\BibitemShut {NoStop}%
\bibitem [{\citenamefont {Shen}\ \emph {et~al.}(2023)\citenamefont {Shen},
  \citenamefont {Barghathi}, \citenamefont {Yu}, \citenamefont {Del~Maestro},\
  and\ \citenamefont {Rubenstein}}]{shen2023stable}%
  \BibitemOpen
  \bibfield  {author} {\bibinfo {author} {\bibfnamefont {T.}~\bibnamefont
  {Shen}}, \bibinfo {author} {\bibfnamefont {H.}~\bibnamefont {Barghathi}},
  \bibinfo {author} {\bibfnamefont {J.}~\bibnamefont {Yu}}, \bibinfo {author}
  {\bibfnamefont {A.}~\bibnamefont {Del~Maestro}},\ and\ \bibinfo {author}
  {\bibfnamefont {B.~M.}\ \bibnamefont {Rubenstein}},\ }\bibfield  {title}
  {\bibinfo {title} {Stable recursive auxiliary field quantum monte carlo
  algorithm in the canonical ensemble: Applications to thermometry and the
  hubbard model},\ }\href@noop {} {\bibfield  {journal} {\bibinfo  {journal}
  {Physical Review E}\ }\textbf {\bibinfo {volume} {107}},\ \bibinfo {pages}
  {055302} (\bibinfo {year} {2023})}\BibitemShut {NoStop}%
\bibitem [{\citenamefont {Dornheim}\ \emph
  {et~al.}(2025{\natexlab{c}})\citenamefont {Dornheim}, \citenamefont
  {Svensson}, \citenamefont {Hamann}, \citenamefont {Schwalbe}, \citenamefont
  {Moldabekov}, \citenamefont {Tolias},\ and\ \citenamefont
  {Vorberger}}]{dornheim2025reweighting}%
  \BibitemOpen
  \bibfield  {author} {\bibinfo {author} {\bibfnamefont {T.}~\bibnamefont
  {Dornheim}}, \bibinfo {author} {\bibfnamefont {P.}~\bibnamefont {Svensson}},
  \bibinfo {author} {\bibfnamefont {P.}~\bibnamefont {Hamann}}, \bibinfo
  {author} {\bibfnamefont {S.}~\bibnamefont {Schwalbe}}, \bibinfo {author}
  {\bibfnamefont {Z.~A.}\ \bibnamefont {Moldabekov}}, \bibinfo {author}
  {\bibfnamefont {P.}~\bibnamefont {Tolias}},\ and\ \bibinfo {author}
  {\bibfnamefont {J.}~\bibnamefont {Vorberger}},\ }\bibfield  {title} {\bibinfo
  {title} {Reweighting estimator for ab initio path integral monte carlo
  simulations of fictitious identical particles},\ }\href@noop {} {\bibfield
  {journal} {\bibinfo  {journal} {The Journal of Chemical Physics}\ }\textbf
  {\bibinfo {volume} {163}},\ \bibinfo {pages} {154101} (\bibinfo {year}
  {2025}{\natexlab{c}})}\BibitemShut {NoStop}%
\bibitem [{\citenamefont {Svensson}\ \emph {et~al.}(2025)\citenamefont
  {Svensson}, \citenamefont {Kalkavouras}, \citenamefont {Hernandez~Acosta},
  \citenamefont {Moldabekov}, \citenamefont {Tolias}, \citenamefont
  {Vorberger},\ and\ \citenamefont {Dornheim}}]{svensson2025accelerated}%
  \BibitemOpen
  \bibfield  {author} {\bibinfo {author} {\bibfnamefont {P.}~\bibnamefont
  {Svensson}}, \bibinfo {author} {\bibfnamefont {F.}~\bibnamefont
  {Kalkavouras}}, \bibinfo {author} {\bibfnamefont {U.}~\bibnamefont
  {Hernandez~Acosta}}, \bibinfo {author} {\bibfnamefont {Z.~A.}\ \bibnamefont
  {Moldabekov}}, \bibinfo {author} {\bibfnamefont {P.}~\bibnamefont {Tolias}},
  \bibinfo {author} {\bibfnamefont {J.}~\bibnamefont {Vorberger}},\ and\
  \bibinfo {author} {\bibfnamefont {T.}~\bibnamefont {Dornheim}},\ }\bibfield
  {title} {\bibinfo {title} {Accelerated free energy estimation in ab initio
  path integral monte carlo simulations},\ }\href@noop {} {\bibfield  {journal}
  {\bibinfo  {journal} {The Journal of Physical Chemistry Letters}\ }\textbf
  {\bibinfo {volume} {16}},\ \bibinfo {pages} {10639} (\bibinfo {year}
  {2025})}\BibitemShut {NoStop}%
\bibitem [{\citenamefont {Dornheim}\ \emph
  {et~al.}(2025{\natexlab{d}})\citenamefont {Dornheim}, \citenamefont
  {Moldabekov}, \citenamefont {Schwalbe}, \citenamefont {Tolias},\ and\
  \citenamefont {Vorberger}}]{dornheim2025fermionic}%
  \BibitemOpen
  \bibfield  {author} {\bibinfo {author} {\bibfnamefont {T.}~\bibnamefont
  {Dornheim}}, \bibinfo {author} {\bibfnamefont {Z.}~\bibnamefont
  {Moldabekov}}, \bibinfo {author} {\bibfnamefont {S.}~\bibnamefont
  {Schwalbe}}, \bibinfo {author} {\bibfnamefont {P.}~\bibnamefont {Tolias}},\
  and\ \bibinfo {author} {\bibfnamefont {J.}~\bibnamefont {Vorberger}},\
  }\bibfield  {title} {\bibinfo {title} {Fermionic free energies from
  \textit{ab initio} path integral monte carlo simulations of fictitious
  identical particles},\ }\href@noop {} {\bibfield  {journal} {\bibinfo
  {journal} {Journal of Chemical Theory and Computation}\ }\textbf {\bibinfo
  {volume} {21}},\ \bibinfo {pages} {7290} (\bibinfo {year}
  {2025}{\natexlab{d}})}\BibitemShut {NoStop}%
\bibitem [{\citenamefont {Ceperley}(1992)}]{ceperley1992path}%
  \BibitemOpen
  \bibfield  {author} {\bibinfo {author} {\bibfnamefont {D.~M.}\ \bibnamefont
  {Ceperley}},\ }\bibfield  {title} {\bibinfo {title} {Path-integral
  calculations of normal liquid {$\text{He}^3$}},\ }\href@noop {} {\bibfield
  {journal} {\bibinfo  {journal} {Physical Review Letters}\ }\textbf {\bibinfo
  {volume} {69}},\ \bibinfo {pages} {331} (\bibinfo {year} {1992})}\BibitemShut
  {NoStop}%
\end{thebibliography}%

\end{document}